\documentclass[1p,12pt]{elsarticle}

\usepackage{amssymb}

\usepackage{amsmath,newtxtext,newtxmath}
\usepackage[scaled=0.86]{helvet}
\usepackage[bookmarks]{hyperref}

\usepackage{color}
\definecolor{blue2}{rgb}{0.36, 0.54, 0.66}
\definecolor{gray}{rgb}{0.5, 0.5, 0.5}
\definecolor{forestgreen}{rgb}{0.13,0.55,0.13}

\usepackage{lipsum}
 
\begin{document}

\title{{Cooperation in Microbial Populations:} \\
{\large Theory and Experimental Model Systems}}

\author[add_jc]{J. Cremer}
\author[add_ef]{A.  Melbinger}
\author[add_ef]{K. Wienand}
\author[add_hj]{T. Henriquez} 
\author[add_hj]{H. Jung\corref{cor}}
\ead{hjung@lmu.de}
\author[add_ef]{E. Frey\corref{cor}}
\ead{frey@lmu.de}

\cortext[cor]{Please address correspondence to Heinrich Jung or Erwin Frey.}

\address[add_jc]{Department of Molecular Immunology and Microbiology, Groningen Biomolecular Sciences  and Biotechnology Institute, University of Groningen, 9747 AG Groningen, The Netherlands.}

\address[add_hj]{Microbiology, Department Biology 1, Ludwig-Maximilians-Universit\"at M\"unchen, Grosshaderner Strasse 2-4, Martinsried, Germany.}

\address[add_ef]{Arnold-Sommerfeld-Center for Theoretical Physics and Center for Nanoscience, Ludwig-Maximilians-Universit\"at M\"unchen, Theresienstrasse 37, D-80333 Munich, Germany.}

\begin{abstract}

{\small

Cooperative behavior, the costly provision of benefits to others, is common across all domains of life. This review article discusses cooperative behavior in the microbial world, mediated by the exchange of extracellular products called public goods. We focus on model species for which the production of a public good and the related growth disadvantage for the producing cells are well described. To unveil the biological and ecological factors promoting the emergence and stability of cooperative traits we take an  interdisciplinary perspective and review insights gained from both mathematical models and well-controlled experimental model systems.  Ecologically, we include  crucial aspects of the microbial life cycle into our analysis and particularly consider population structures where an ensemble of local communities (sub populations) continuously emerge, grow, and disappear again. Biologically, we explicitly consider the synthesis and regulation of public good production. The discussion of the theoretical approaches includes  general evolutionary concepts, population dynamics, and evolutionary game theory. As a specific but generic biological example we consider populations of \textit{Pseudomonas putida} and its regulation and utilization of pyoverdines, iron scavenging molecules. The review closes with an overview on cooperation in spatially extended systems and also provides a critical assessment of the insights gained from the experimental and  theoretical studies discussed. Current challenges and important new research opportunities are discussed, including the biochemical regulation of public goods, more realistic ecological scenarios resembling native environments, cell to cell signalling, and multi-species communities.

}
\end{abstract}

 
\maketitle

\cleardoublepage\newpage
	
\tableofcontents

\cleardoublepage\newpage

\section{Introduction}
\label{sec:introduction}

\subsection{The conundrum of cooperative behavior in the theory of evolution}
\label{sec:introconundrum}

Cooperative behavior in human societies is defined as an interaction between individuals directed towards a common goal that is mutually beneficial. 
Such `social' behavior is not restricted to humans but actually widespread in nature. Variants of it can be found in animal populations, down to insect societies and even microbial populations.
How can one reconcile such `altruistic' behavior with the fact that organisms are generally perceived as being inherently competitive? Addressing this conundrum 
in evolutionary biology, Darwin wrote
in his book ``Origin of the Species''~\cite{Darwin}:

\begin{quote}
\textit{
"... one special difficulty, which at first appeared to me insuperable, and actually fatal to my whole theory."\footnote{The citation actually refers to eusociality, which is special kind of cooperative behavior, found, for example, in some insect populations.} 
}
\end{quote}

What difficulty exactly is Darwin referring to? 
A key element in the Darwinian theory of evolution is \textit{natural selection}, i.e.\/ the differential survival and reproduction of individuals in a population that exhibit different traits (including different types of behavior). Cooperative behavior, while beneficial to all or some other individuals present in a population, is costly to individuals exhibiting that trait. This entails a fitness disadvantage which ultimately should lead to the extinction of all the individuals that exhibit cooperative behavior.
How, despite this, cooperation is maintained --- or has evolved in the first place --- is a conundrum in evolutionary biology~\cite{Dawkins, Axelrod:1984}. 

The puzzling aspect of cooperative behavior can also be illustrated by comparing its benefits at different \textit{levels of a population}: 
A population (or a society) \textit{as a whole} might benefit from the cooperative behavior of a subset of individuals.
However, at an \textit{individual level}, this behavior can be `exploited' by other individuals (often called `free-riders') that benefit from the cooperative behavior but do not participate in such cooperative behavior.
As a consequence, cooperating individuals die out to everyone's loss. This circumstance is known as the \emph{`dilemma of cooperation'}. 
As we will learn in the following, this dilemma is actually part of a possible answer to Darwin's difficulty, as it hints towards the importance of \textit{population structure and spatial organization} in a population for the evolution and maintenance of cooperation and biological diversity in general.

\subsection{The dilemma of cooperation in the microbial world}
\label{sec:subseccooperationinthemicrobialworld}

A myriad of theoretical and experimental studies have investigated different aspects of cooperative traits and a broad variety of mechanisms ensuring their emergence and evolutionary stability, covering a variety of biological systems, reviewed in Refs.~\cite{Keller:1999, Kerr:2004gx, Doebeli:2005, Nowak:2006, West:2007, Frey:2010, Archetti-Review:2012, Sanchez_Gore:2013}. 
In this review, we will focus on cooperative behavior in the microbial world:
Bacteria and other microbial cells mostly live in communities, often consisting of multiple phenotypes or species~\cite{Konopka:2009kl,
Flemming:2019, Cordovez:2019, Costea:2018, Mukherjee:2019}. 
Interactions between individuals in these systems are typically mediated by the secretion of various kinds of extracellular products (exoproducts) including metabolites, exoenzymes like siderophores, matrix components in biofilms, signalling molecules, and different types of toxins~\cite{Wolin:1997jz,Seth:2014,Abreu:2016,Barth373,Sutherland:2001kp}.
A particular kind of exoproducts are those that benefit others in a community, which in the following we will refer to as ``\textit{public goods}". If, in addition, the synthesis of a public good is costly to the producing cells, it is commonly referred to as \textit{cooperative behavior}~\cite{Wingreen:2006, West_Diggle:2007, Frey:2010, Allen:2013kl, Damore_Gore:2012, Ozkaya:2017dl, Cavaliere:2017fh, Smith:2019hp}. In our discussions we will focus on systems where genetic differences are small and linked to the public good production. We will not discuss other genetic changes beyond these, like fundamental changes in metabolism or other fundamental physiological processes common in multi-species communities.



Our objective is to address the question of emergence and stability of cooperative traits within microbial populations from an interdisciplinary perspective that discusses both, \emph{abstract mathematical models} which originated from evolutionary biology), and \emph{experimental model systems} where specific microbial species are studied under well-controlled laboratory conditions. This interdisciplinary perspective will require from a reader with either background to show some willingness to learn about the respective other field, and we hope to provide sufficient details to guide readers from either fields.

The rapid advancement of research on microbial communities and cooperation makes it necessary to further confine the range of topics
that we address in this review article.
We restrict ourselves to the discussion of well-characterized bacterial systems much simpler than the complex community structures native microbial populations on this planet might show~\cite{Nemergut:2013, Stubbendieck:2016hv, Costea:2018, Cordovez:2019}.
Further, in describing the theoretical work in the field, we will mainly discuss the work that links to these systems and discuss their relation to more general approaches investigating cooperation. 
Lastly, we will mostly confine ourselves to locally well-mixed scenarios, for which spatial effects can be neglected. In particular, we do not consider the rich spatial arrangement of microbes within colonies and biofilms~\cite{Nadell:2016, Flemming:2019}. Explicitly accounting for space leads to a plethora of intriguing and important phenomena and we give a short (but incomplete) review of recent progress in the field towards the end of this review article.
While we attempt to provide a broader overview of the field, the specific examples discussed in detail follow our personal research background, and we apologize for not covering other important work on microbial cooperation.

Given the ecological variety of microbial life and the biochemical complexity of cells and their interactions, many different aspects can be important in shaping the evolution of microbial populations and the emergence and stability of their (cooperative) traits. Throughout this review, we repeatedly discuss three such aspects in great detail, which we think are particularly important to consider:

\begin{itemize}
\item Microbial populations are highly structured: Evolutionary dynamics is occurring simultaneously in a set of different sub-populations and these sub-populations continuously emerge and disappear over time.
\item The life cycles of microbes include strong phases of growth, and sub-populations can vary in size over several orders of magnitude --- from a few initial cells (if not a single one) to the billions or even trillions of an established community. 
\item Public goods are not simply continuously produced by the cells. Instead, as it happens for many other phenotypes, regulatory networks tightly control the expression of public goods, based on other cellular processes and the environmental conditions cells sense.
\end{itemize}

A proper consideration of public goods, or  --- more generally --- exoproducts in bacteria thus requires the consideration of population structure, growth, stochastic effects of demographic and environmental noise, as well as the biological aspects of public good synthesis and utilization.
The aspects of population structure and growth dynamics across bacterial life-cycles can already be considered by theoretical considerations alone and we discuss those and their relations to formulations of evolutionary theory. 
While equally important, the regulatory aspects of public good production require a detailed consideration of the specific biological system one aims to understand. 
In this review, we will illustrate this requirement by considering specifically the bacteria \emph{Pseudomonas putida}  and its regulation and utilization of pyoverdines, which are public goods produced to support iron uptake. 

\subsection{Outline} 
\label{sec:sub_outline}

In the following Section \ref{sec:cooperationinmicrobialworld} we will first discuss the dilemma of cooperation from a broader perspective that also includes cooperation of higher organisms. 
This is mainly meant to give the reader some background on the long and convoluted history of the topic (\ref{sec:intro_dilemma_cooperation}). 
Next, we will discuss the `prisoner's dilemma' (\ref{sec:intro_game_theory}), a classical example that illustrates the dilemma of cooperation. 
This is followed by a concise review of the possible conceptual resolutions of the dilemma: assortment and reciprocity (\ref{Reciprocityandassortment}). 

Section \ref{sec:microbial_communities} discusses important characteristics of microbial life and evolution. It provides an overview of the literature on experimental model systems studying cooperative interactions in bacterial populations (\ref{sec:growth_in_dynamicallyrestructuringpopulations}), and discusses assortment as an essential factor for a resolution of the cooperation dilemma (\ref{sec:factors_cooperationinmicrobialcooperation}). 
This is followed by a short introduction into microbial model systems in Section~\ref{sec:experimental_model_systems} which are used to study evolution (\ref{sec:evolution_mircrobes}), cooperation via public good production (\ref{sec:public_good_microbes_systemstostudy}), and the role of structured populations (\ref{sec:microbes_artificial_enviroenments}) in laboratory populations.

In Section \ref{sec:mathematical_formulation}, we will then review  the most important theoretical concepts and mathematical methods available to consider evolutionary dynamics and the dilemma of cooperation in well-mixed populations.
Concepts discussed include the Price and replicator equations (\ref{sec:price_equation} and \ref{sec:replicator_equation}), evolutionary game theory (\ref{sec:evolutionary_game_theory}), and the theory of stochastic processes (\ref{sec:fluctuations}). 
We will also discuss models of population dynamics to specifically consider growth of microbial populations (\ref{sec:popgrowth}).  
We will keep the level of mathematical detail to a minimum and focus mainly on those aspects of the theories that are of key relevance for the following discussions.

Section~\ref{sec:evolinstructuredkinandgroup}  introduces the general concepts available to consider evolution in structured populations. The concepts of group- and kin-selection are critically discussed (\ref{sec:group_selection} and \ref{kin_selection}), as are the two-level consideration based on the Price equation and Hamilton's rule (\ref{sec:twolevelselectionandhamiltonsrule} and \ref{sec:note_hamiltonsrul}). 
This is merely to explain these often-used and historically loaded concepts in the context of this review.

Section \ref{sec:pyoverdine} provides then a detailed overview of the biology of the public good pyoverdine  in \textit{Pseudomonas} populations. 
It includes the biochemical characterization of pyoverdine production and regulation to illustrate the considerable complexity of the public good production in microbial systems. 
Moreover, we
discuss why \textit{Pseudomonas putida} can serve as a well-defined experimental model system for studying cooperation in bacterial populations.

This experimental characterization of a specific biological system is then the basis for the mathematical models discussed in Section \ref{sec:selection_drift}, where we  elucidate  the evolutionary dynamics of cooperative behavior in structured populations, focusing on pyoverdine production as an example. 
The section reviews recent advances in understanding the maintenance and evolution of cooperation in bacterial populations that have a life-cycle populations structure. 
The discussion shows how experiment and theory complement each other to dissect the role of environmental noise, demographic noise and selection. In the first Section \ref{sec:random_drift_bact_pop}, we will illustrate the role of assortment noise in growing bacterial populations in which the selection pressure is weak. 
This is followed by a theoretical analysis of the combined role of selection pressure, growth advantage of more cooperative sub-populations, and demographic noise on the dynamics of an ensemble of populations containing cooperators and defectors (\ref{sec:egt_growing_populations}).
The main insight will be that the interplay between these factors can lead to the emergence of a transient increase in cooperator fraction in the whole population. 
Next, we review experiments and detailed mathematical modeling of a \textit{P. putida} model system that confirms these predictions qualitatively and elude on the role of molecular features of public goods (\ref{sec:explicit_public_good}).
Combining the above, we discuss how life cycles can lead to both the maintenance and the evolution of cooperation in bacterial populations (\ref{sec:life_cycles}).

Finally, we will briefly review spatially extended systems in Section~\ref{sec:spatially_extended_systems}, and conclude with a concise summary and a brief outlook in Section~\ref{sec:conclusions_outlook}.

\section{The dilemma of cooperation and possible resolutions}
\label{sec:cooperationinmicrobialworld}

\subsection{A broader perspective of the dilemma of cooperation} 
\label{sec:intro_dilemma_cooperation}

Before focusing on bacterial systems we would like to broaden our perspective for a moment and discuss cooperation in general terms elucidating the variety and omnipresence of such social behaviour; this summary follows Ref.~\cite{Cremer:2011thesis}.
In human behaviour, cooperation and the ensuing dilemma can be found on almost every interaction scale and in diverse fields ranging from psychology, to sociology, politics, and  economics. 
This starts with interactions of individuals in small entities like families and ordinary tasks like sharing responsibilities in a household and goes to humanity as a whole, for instance in facing the global challenge of climate change. 
Humans are endowed with a broad range of mechanisms promoting cooperation~\cite{Fehr:2003}: Due to our ability to recognise and remember other individuals, we can (to a certain extend) distinguish cooperators from cheaters and thereby prevent interactions with cheaters, warn others, or even punish cheaters~\cite{Boyd:2003}. 
Nevertheless, different plots of the dilemma of cooperation are still omnipresent in human life. 
Hence, the origin and nature of cooperativity in human societies and its limitations is a heavily debated topic~\cite{Tomasello, Sober:1998}: 
Is cooperative behavior an inherent characteristic of human beings?  
How important are early childhood experiences and the first social interactions with other humans? How does culture come into play? 
Which role does punishment and the ability to form institutions have? 
Which aspects are special for humans and in which respect does the cooperative behavior of \emph{Homo sapiens} differ from other \emph{Hominidae}? 

Beyond humans, cooperation is also widespread in the animal kingdom. 
Examples include the herd formation of gregarious animals~\cite{Clutton-Brock:2009}. 
While beneficial for the whole population, animals on the outer edges take a higher risk of predation. 
Executing alarm calls, as observed for birds and monkeys, is another strong form of cooperation~\cite{Clutton-Brock:2009,Sober:1998}. 
The surrounding individuals are warned, while at the same time the caller is strongly increasing the attention of the discovered predator.

Another often stated extreme form of cooperation is the separation of working and reproducing individuals in insect populations, see e.g.\/ Ref.~\cite{Holldobler:2009,Gordon:2015jd}. 
Why for example are most of the individuals sterile female `workers' or other specialized individuals supporting the reproduction of one or a few fertile `queens'?
The classical explanation for cooperation within such colonies or super-organisms is the strong relatedness of kin~\cite{NowakWilson}.
From a 'gene's eye view', genetically identical workers still reproduce their genes by supporting the queen. 
However, the precise reasons for cooperation in insect colonies and protective measures against genetically different individuals are  more subtle, and different species might adopt different mechanisms~\cite{Keller:1999, Ratnieks:2006, Libbrecht:2013}. This includes kin discrimination and reciprocity~\cite{Foster:2006}.

Finally, for unicellular organisms cooperation is widespread as well. 
As mentioned already in the introduction, microbial populations cooperation is often given by the production of a public good~
~\cite{Velicer:2003p377, Kreft:2005, West:2006, Brockhurst:2007, Gardner, Gore:2009, Hallatschek:2011, Buckling:2007}.
Striking examples include the synthesis of matrix-proteins for biofilm formation, or the production of extracellularly acting enzymes for better nutrient or mineral uptake~\cite{Flemming:2019, Nadell:2016, Gore:2009}.
Another well-studied example of cooperation in microbes is the formation of fruiting bodies, for example in the slime mold \emph{Dictyostelium discoideum}~\cite{Strassmann:2000, Santorelli:2008}.  
While formation increases dispersal rates and therefore the exploitation of new nutrient resources, cooperation involves altruism as stalk cells cannot disperse but die. 

Mechanisms described to maintain cooperation involve, for example, limited diffusion of a public good, spatial restrictions and cell-cell contacts~\cite{Kummerli_Gardner:2009, Kummerli_Griffin:2009,Julou:2013}, metabolic constraints controlling social cheating~\cite{Dandekar:2012}, or the presence of a loner strain in a producer and non-producer system~\cite{inglis_presence_2016}. We discuss a few examples of microbial cooperation in more detail in Section (\ref{sec:evolution_mircrobes}), but first consider the prisoner's dilemma, the classical and most famous example of game theory, to illustrate the dilemma of cooperative behavior.

\subsection{Game theory: the prisoner's dilemma and public good games}
\label{sec:intro_game_theory}

To further illustrate the dilemma of cooperation, let us consider one specific situation, the prisoner's dilemma~\cite{Axelrod:1984}, which has become a mathematical metaphor to describe cooperative behavior~\cite{Maynard, Nowak:2006, Nowakbook, Frey:2010}. The original formulation refers to a scenario where two criminals are interrogated. Each criminal can testify against the other (non-cooperating behavior) or remain silent (cooperating behavior)~\cite{Axelrod:1984}. Here, we present it as \emph{public good game} where individuals adopting two different strategies, called `cooperation' and `defection', play against each other. 
The pairwise interactions between these two different `strategies' are summarised in what is called a payoff matrix:
\begin{displaymath}
\label{eq:payoff}
\begin{array}{l|ll}
\mathbf{P} & \text{cooperator} \, 
           & \text{defector} \,
\\ \hline
\text{cooperator} & b-c & -c   \\
\text{defector} & b   & 0  
\end{array}
\end{displaymath}
A `cooperator' provides a benefit $b$ at a cost $c$ to itself (with $b-c>0$). 
In contrast, a `defector' (or `free-rider') `refuses' to provide any benefit and hence does not pay any costs. 
Thus, in an interacting between two cooperators, both obtain the effective payoff $b-c$. 
If a cooperator interacts with a defector, the defector obtains the benefit $b$, while the cooperator does not obtain any benefit but still has to pay the costs $c$ (negative payoff $-c$). 
In an interaction between two defectors there are no costs but also no benefits.
Hence, for the `selfish' individual (`defector'), irrespective of whether the competitor cooperates or defects, defection is always favourable, as it avoids the cost of cooperation, exploits cooperators, and ensures not to become exploited.
In other words, adopting the strategy `defection' is the only strategy that can not be exploited; in game theory it is called a Nash-equilibrium~\cite{Maynard}. 
The dilemma is that everybody is then, with a gain of $0$, worse off compared to a state of universal cooperation, where a net gain of $b-c>0$ would be achieved.

Instead of viewing the public good game as a strategic game, it can also be interpreted as a population dynamics problem where individuals interact in a pairwise fashion with a reproductive fitness determined by the payoff matrix~\cite{Maynard}; for a mathematical formulation see Section~\ref{sec:evolutionary_game_theory}.
As defectors are always better off in pairwise interactions with cooperators, their number will increase in the population such that in the long run there will only be defectors. 
Mathematically, this can be formulated as a differential equation for the fraction of cooperators, $x$; for more details see later in Section~\ref{sec:evolutionary_game_theory}.
Assuming that every individual interacts with all other individuals in a population with equal probability, and taking the expected payoff-values as expected fitness-values, the temporal change in the fraction of cooperators, $x$, follows the equation
\begin{equation}
	\frac{dx}{dt} 
	= 
	-\, c \, x \, (1-x) \, .
\label{eq:replicator_pd}
\end{equation}
For $c>0$, and independent of the initial amount of cooperators the dynamics always declines towards the state $x=0$ (no cooperators), which is hence called an attractive fixed point of the dynamics.
In this sense, the above Nash equilibrium is also called evolutionary stable. Not cooperating is an `evolutionary stable strategy'~\cite{Maynard:1995}.

The prisoner's dilemma in its evolutionary formulation is a paradigmatic example in evolutionary game theory, a theoretical framework considered in more detail in Section \ref{sec:evolutionary_game_theory}. 
It nicely shows how fitness can be motivated heuristically without reference to a specific biological systems and further illustrates the dilemma of cooperation.
However, it should not be mistaken as a realistic model for an actual biological or ecological process like the cooperative dynamics within a bacterial population. In real systems, interactions between individuals are much more complex and there are many important biological or ecological factors which need to be considered. For microbial systems, we have already mentioned growth in structured populations and the regulatory control of public goods as important aspects and we discuss these and others in more detail in the following Section \ref{sec:microbial_communities}.  An in-depth discussion of realistic cost and benefit functions is provided in Section~\ref{sec:pyoverdine} for the example of pyoverdine producing bacteria.

\subsection{Reciprocity and assortment can stabilize cooperation}
\label{Reciprocityandassortment}

In view of the complexity of biological systems and the manifold types of cooperative behaviour it would be surprising to find a universal answer to the questions how cooperative behavior originated and how is it maintained.
In fact, the solutions to the cooperation dilemma are as diverse as the observed forms of cooperation~\cite{Trivers:1971, Sober:1998, Keller:1999, Hibbing:2010, Celiker2013, Keller:1999, Kerr:2004gx, Nowak:2006, West:2007, Frey:2010, Sanchez_Gore:2013, Archetti-Review:2012}. However, at a conceptual level, one can roughly distinguish between two main classes of mechanisms: reciprocity and assortment.\footnote{Note that this classification is not unique and other authors might prefer to sort using different categories. \cite{Keller:1999, Kerr:2004gx, Zwietering:1990, Nowak:2006, West:2007}.}

\paragraph{Reciprocity} 
If individuals have sophisticated skills like the ability to  recognise other individuals and  memorise their behaviour, they might actively adjust their behavior to obstruct the exploitation of non-cooperators to themselves or others: cooperation is maintained by reciprocity \cite{Trivers:1971, Nowak:2006, Sober:1998}. 
In general, one distinguishes between direct and indirect reciprocity. 
Direct reciprocity builds on repeated interactions. For example, in the repeated prisoner's dilemma game it includes the famous `tit for tat' strategy \cite{Axelrod:1984}, where individuals continue cooperating only if playing with another individual that was cooperating during the last engagement. 
Indirect reciprocity also accounts for third parties and some sort of communication. More complex forms of memory-based mechanisms promoting cooperation include punishment \cite{Boyd:2003, rockenbach-2006-444, Hauert_Traulsen:2007, Okasha, Nowak:2006} and policing \cite{Fehr:2003, Yamagishi}.

\paragraph{Assortment} 
Cooperation may also be facilitated by a high degree of relatedness among interacting individuals such that cooperators interact more likely with each other than with non-cooperating free-riders. In such a situation cooperators can benefit from other cooperators, and they also run a lower risk to be exploited by non-cooperating free-riders. Possible ecological scenarios promoting such an assortment include spatially extended systems~\cite{Nowak_Bonhoeffer:1994, Hauert:2004, Roca:2009, Gelimson:2013, Bauer_Frey_EPL:2018, Bauer_Frey_PRE:2018, Bauer_Frey_PRL:2018}, populations structured into distinct sub-populations~\cite{Wilson:1977a, Silva:1999, Fletcher:2004, Kreft:2005, Janssen:2006,  Killingback:2006, Michod:2006,Tarnita:2009a, Tarnita:2009, MacLean:2010} or more complex interactions between different individuals within a population (networks)~\cite{Lieberman:2005, Santos:2006, Ohtsuki:2006, Pacheco, Altrock:2017kb}.

\section{Microbial communities} 
\label{sec:microbial_communities}

Microbes are the most widespread life form on our planet. 
These organisms, which appear simple only at first glance, exhibit tremendous diversity and are able to adapt to a multitude of changing environmental conditions \cite{Brooks:2011,Sauer:2002, Wood:2011, Jung:2019}.
For example, they balance the cellular demands to optimize growth, uptake and survival under a range of environmental conditions by changing the composition of expressed proteins~\cite{Wood:2011, Bleuven:2016, Hui:2015ig}. But strategies to adapt do not only involve a controlled change of protein resources. Cells also control the conservation and change of their own DNA integrity. For instance, during prolonged starvation \textit{Bacillus subtilis} differentiates into phenotypic distinct cell types to realize different survival strategies like the uptake of foreign DNA (competence)~\cite{Finkel1999,  Chen2004, Leisner:2008, Leisner:2009}. Furthermore, microbes often live in complex communities where they interact in various ways. Besides the already mentioned production of public goods, this includes communication via signaling like quorum sensing~\cite{Bassler:1999hy, Miller:2001, Nadell:2008, Bauer_Knebel:2017, Kaur:2018, Xavier:2018, Tobias:2019, Dandekar:2012, Wang:2015, Jung:2011}, the exchange of metabolites~\cite{Gore:2009,Nemergut:2013, Estrela:2016,Cremer:2016dz,Lustri:2017, Friedman:2017bz,DSouza:2018}, but also competition for nutrients and the accumulation of waste-products~\cite{Freilich:2011, Nadell:2011, Nadell:2013, Schluter:2015hi, Frost:2018ec, Goldford:2018jf,Chacon:2018, Warren:2019cl}. 
These interactions are often occurring within dense biofilm communities~\cite{Stoodley:2002, Sauer:2002, Stewart:2008,  Drescher:2014, Nadell:2016}. 

The realization of the existence of these manifold interactions has led to the establishment of microbial community research as an important field of  microbiology. Cooperation via public goods, exchange of metabolites, and  signalling molecules in these communities is often so pronounced that some researches even see microbial communities  as social entities performing sophisticated processes like division of labor and communication, although anthropomorphic wording should be chosen with care~\cite{Velicer:2003p377, Kolter:2006, West:2006, West_Diggle:2007, Gordon:2015jd, Dragos:2018}. 
As many microbial communities, such as many biofilms, are spatially heterogeneous entities built up of many independent subunits, they have been even posited by some authors as model systems for understanding the evolution of multi-cellularity~\cite{Rainey:2010ea,Nadell:2009,Ratcliff:2012hy}.

\subsection{Assortment in microbial populations}
\label{sec:factors_cooperationinmicrobialcooperation}
Given these variant microbial lifestyles, what are possible mechanisms and principles promoting cooperative behavior? 
Since microbial organisms are limited in their ability to recognise specific other individuals and memorise their behavior, they can hardly rely on reciprocity based mechanisms to ensure cooperation. Correspondingly, assortment mechanism are crucial to overcome the dilemma of cooperation. For microbial populations, assortment can be facilitated by a number of biological and ecological factors. As for assortment mechanisms for other organisms, the factors can roughly be divided into two classes:
\emph{active assortment} and \emph{passive assortment}.\footnote{As for the classification of the more general mechanisms promoting cooperation, other authors might prefer other classification schemes.}

\paragraph{Active assortment}
For active assortment, individuals themselves contribute actively to a positive assortment; cooperators `preferentially' engage with other cooperators. While this does not require the ability to memorise previous interactions, it still requries the capability of cooperators to identify other cooperators. While such a form of kin-discrimination might be present in higher developed organisms, like in animal societies~\cite{Holldobler:2009}, examples in less sophisticated forms of life have also been proposed. This includes in particular the idea of `green-beard' genes \cite{Hamilton:1963, Gardner:2010, West:2010,Queller:2003, Smukalla:2008} which  directly encode for cooperative behavior and also some recognition mechanism, allowing for cooperative individuals to actively recognize the cooperative trait of others. However, the stable realization of such green-beard mechanisms might be limited in reality as cheating mutants which simply pretend to be cooperators can emerge~\cite{Gardner:2010, West:2010, Jansen_Minus:2006, Traulsen:2007}. 

\paragraph{Passive assortment}
Accordingly, passive assortment is likely the predominant scenario to stabilize cooperation in microbial populations. Cooperating microbes  engage more often with other cooperators due to the external environmental conditions. One scenario of such passive assortment has already been suggested by Hamilton, who argued that limited dispersal and mixing can lead to the coexistence of related individuals (like cooperators) close to each other~\cite{Hamilton:1964}. For example, a high viscosity of the surrounding media can hinder motility and hence cooperating individuals might more likely `interact' with other neighboring cooperating individuals. An extreme form of this scenario are spatially extended populations where through spatial clustering (immobile) cooperators preferentially interact with other cooperators~\cite{Hamilton:1964, Nowak_Bonhoeffer:1994, Roca:2009a,  Durrett:1994}.

\subsection{Growth and dynamically restructuring populations} 
\label{sec:growth_in_dynamicallyrestructuringpopulations}

Given the diversity of microbial communities and their habitats, the ecological and biological factors shaping assortment dynamics in microbial populations can be manifold, varying tremendously from species to species and from habitat to habitat. However, as we mentioned already in the introduction, a few factors appear to be very generic characteristics of microbial life and important to consider assortment dynamics and evolution.

\paragraph{Microbial growth} The first aspect important to consider is microbial growth: Many microbial organisms can grow very fast, provided they encounter the right conditions. Doubling times can be as low as 20 minutes~\cite{Monod:1949, Neidhardt:1999, Gibson:2018}, and this fast growth is important for bacterial cells to compete with other species within their microbial community~\cite{Khatri:2012,Merritt:2018,Cremer:1LtZBIVB}. 
The importance of growth for the fitness of microbial cells is emphasized by the carefully coordinated regulation of genes. Consider for example the allocation of protein-synthesis resources into synthesis-related and other proteins: the condition-dependent regulation of ribosomes ensures their optimal utilization, preventing their waste-full and growth-limiting over-expression~\cite{Scott:2010cx, Scott:2014jo, Hui:2015ig, Basan:2015eu, Zwietering:1990, Hermsen:2015, Towbin:2017ca}. 
Fast growth and nutrient availability is also the basis for the huge number of microbial cells observed in many habitats. As the fast growth of microbes relies on the supply of nutrients, the abundance of microbial cells is tied to the availability of nutrient sources. Thus, as a consequence of distributed nutrient sources continuously changing over space and time, microbial populations are typically highly structured. Many different sub-populations form a population~\cite{Hall-Stoodley:2004, West:2006, Flemming:2019, Klauck:2018}.
The exact structure depends on the ecological specifics a population encounters, but many examples can illustrate the structuring. 
For instance, populations of gut-bacteria are distributed across many hosts, and within the gut digestible food particles,  are heterogeneously distributed~\cite{Capuano:2016gz, Cremer:2017bm, Arnoldini:2018cc}. 
Marine bacteria specialized in degrading organic matter occupy zillion flakes of marine snow~\cite{Oceanography:vi, Kiorboe:2003, Grossart:2006}, and the lab-strain \emph{E. coli} MG1655 is distributed across several laboratories worldwide. 
Evolutionary dynamics is steadily happening simultaneously in all sub-populations.

\paragraph{Dynamic restructuring of populations} A further important aspect of structured populations is the dynamics of their restructuring. Particularly, despite strong growth dynamics and huge cell numbers that can be reached within each sub-population, sizes of the locally confined populations can also be very small. This is especially the case immediately after the occupation of newly available habitats where prolonged phases of growth have not occurred yet and sub-populations can go through narrow bottlenecks. In the extreme limit, growth in newly available habitats start with only a single or a few bacterial cells, dispersing from other sub-populations.
To illustrate this dynamical restructuring process, consider again the different examples mentioned before:
The intestine of the mammalian gut gets occupied starting from birth with a small number of bacteria, for example \textit{Bifidobacterium} cells, intestinal populations of different bacterial species then start to emerge occupying their specific niches over the first few months~\cite{Seedorf:2014ey, LaforestLapointe:2017, Verkhnyatskaya:2019}. 
Related to the nutrient intake of the host, new clusters of nutrient sources, like clusters of resistant starch and fibers, reach the intestine every day, and are captured by bacteria when they reach the distal parts of the intestine. 
A similar restructuring of the microbial population dynamics is happening for marine bacteria feeding on marine snow. These debris particles are continuously supplied from upper layers of the ocean and first need to be occupied by the marine bacteria feeding on them~\cite{Oceanography:vi,Grossart:2006}.
Finally, consider the process of `plating' of bacterial cells in the laboratory as an artificial example. 
Here, bacterial cultures are diluted to low densities such that the spreading of the culture onto a growth supporting agar plate leads to growth of separated microbial colonies, each starting with a single cell. 
For \emph{E. coli} MG1655 repeated plating led to the formation of various laboratory stocks and some substantial variation has occurred across the meta-population~\cite{Hayashi:2006ec}. 


Thus, overall, microbial populations are highly structured and the dynamical restructuring process often involves bottlenecks and phases of strong growth. However, the detailed characteristics of the population structure, growth and the restructuring process illustrated here can vary strongly from example to example and the involved time and length scales might change considerably.

Accordingly, passive forms of assortment in dynamically restructuring populations might  provoke sub-populations where cooperators engage predominantly with other cooperators. However, non-cooperators are always better off than neighbouring cooperating individuals and hence the positive assortment of cooperators persistently has to overcome this direct advantage of the non-cooperators. Given this competition and the variation of assortment processes illustrated above, simply mentioning assortment and the clustering of cooperating individuals alone is not answering how cooperative behavior is maintained. Instead, one has to study the details of assortment dynamics and how they lead to the evolutionary stability of cooperation and public good production.

Adding to complexity,  microbes themselves can also actively influence their life cycle by sensing environmental conditions and reacting to it - a factor which should be explicitely considered for many species when studying cooperation.  For example, studies of \emph{P. aeruginosa}~\cite{Stoodley:2002, Sauer:2002, Hall-Stoodley:2004} suggest that  typical life cycles of biofilm-forming pseudomonads pass through different steps, with dispersal events which are triggered by the local collections of cells and the nutrient conditions they encounter. 

In the remaining part of this review we focus on the case of structured populations with well-mixed sub-populations when studying the emergence and stability of cooperation. We focus on well-characterized bacterial systems and precisely defined synthetic environments. 

\section{Studying evolution and cooperative behavior under controlled laboratory conditions}
\label{sec:experimental_model_systems}

\subsection{Evolutionary studies with microbes in the laboratory}
\label{sec:evolution_mircrobes}

Microbial populations offer unique opportunities to experimentally study evolution: Microbial cells grow fast, leading to large population sizes and evolution on fast time-scales. Moreover, samples can easily be stored and analyzed at later time points~\cite{Jessup:2004ds, Conrad:2011, Elena:2003, Kawecki:2012cm}.
By now microbial systems have been used to study different aspects of evolutionary dynamic in the laboratory~\cite{Good:2018il, Elena:2003}. Examples include long-term evolution experiments~\cite{Elena:2003, Kryazhimskiy:2014, Lenski:2017, Good:2017cu} observing adjustment in fitness over thousands of generations, the evolution of speciation in continues culture~\cite{Rosenzweig:1994,Koeppel:2013}, and the adjustment of swimming and chemotaxis in spatially extended habitats~\cite{Baym:2016hx, Fraebel:2017gw, BinNi:2017, Liu:51NDZHe4}. 

\subsection{Microbial systems to study public goods}
\label{sec:public_good_microbes_systemstostudy}

Since some time now, laboratory experiments have also been used to study cooperative behavior of bacteria and the selection dynamics in different highly controlled environments. 
Much research concerning the cooperative behaviour of bacteria is performed for well-characterized microbes in simplified experimental model systems offering the possibility to gain understanding under well-defined conditions. 
In the following we briefly discuss some of the employed microbial systems, but many more have been studied; see e.g.\/ Refs.~\cite{Buckling:2009, Wingreen:2006, West_Diggle:2007, Allen:2013kl, Damore_Gore:2012,Li:2015bp, Ozkaya:2017dl, Cavaliere:2017fh, McNally:2017}.

An often used model system for a cooperating microbe is the proteobacterium \textit{Pseudomonas aeruginosa}~\cite{Buckling:2007, Harrison_Buckling:2009, Kummerli_Brown:2010, Kummerli:2015, Becker_Wienand:2018}. 
Iron, which is usually bound in large clusters, is essential for the metabolism of these bacteria. 
Therefore, some individuals, the so-called producers, provide  siderophores, which are iron scavenging organic compounds. 
Producers release them as a public good into the environment. 
Because of their large binding affinity to iron, these compounds can solve single iron molecules and build siderophore-iron complexes~ \cite{Kummerli_Brown:2010, Kummerli:2015}. 
The freely diffusing complexes can be equally taken up by cooperators and non-contributing free-riders. 
The dilemma arises due to the metabolic costs associated with the production of the public good: 
Producers replicate slower than free-riders and thereby have a fitness disadvantage. 
We discuss a related iron scavenging system active in \textit{P. putida} in detail in Section~\ref{sec:pyoverdine}.

Another prominent and well-studied example is given by invertases hydrolyzing disaccharides into monosaccharides in the budding yeast \textit{Saccaromyces cerevisiae}~\cite{Greig_Travisano:2004, Gore:2009, MacLean:2010}.
Budding yeast prefers to use the monosaccharides glucose and fructose as carbon sources. 
If they have to grow on sucrose instead, the disaccharide must first be hydrolyzed by the enzyme invertase. 
Since most of the produced monosaccharides ($99 \%$) diffuses away and is shared with neighboring cells, it constitutes a public good available to the whole microbial community. 
This makes the population susceptible to invasion by mutants  that save the metabolic cost of producing invertase. 
Naively, this suggests that yeast is playing the prisoner's dilemma game and strains not producing the public good are able to invade the population leading to the extinction of the producing wild type strain.  
But, this is not the case. 
Gore and collaborators~\cite{Gore:2009} have shown that the dynamics is rather described by a snowdrift game, in which cheating can be profitable, but is not necessarily the best strategy if others are cheating too. 
The underlying reason is that the growth rate as a function of glucose is highly concave and, as a consequence, the fitness function is \emph{non-linear} in the payoffs.
The lesson to be learned from this investigation is that defining a payoff function is not a trivial matter, and a naive replicator dynamics fails to describe biological reality. 
As we will see in Section~\ref{sec:model_system_pseudomonas}, the same is true for \textit{Pseudomonas} populations. 
Hence, we believe that -- quite generally -- it is necessary to have a detailed look on the nature of the biochemical processes responsible for the growth rates of the competing microbes.

Other well-studied examples exist as well. This includes the production of polymers as public goods to hold communities together and provide to their well-being, see e.g.\/ Refs.~\cite{Rainey:2003,Xavier:2007gp, Drescher:2014, Dragos:2018}. 
For example, the expression of sticky polymers by \textit{P. fluorescens} allows for the formation of stable microbial colonies which enable the floating on liquids. Populations might benefit by the effective access to oxygen at the air-liquid interface~\cite{Rainey:2003}. 
Further examples discussed include the metabolic utilization of different carbon sources which require the activity of digestive enzymes released by cells into the environment \cite{Greig_Travisano:2004, Koranda:2013, Rainey:2003, RakoffNahoum:2016}.

\subsection{Studying population structure in artificial enviornments}
\label{sec:microbes_artificial_enviroenments}

To consider the influence of population structure on evolution and the stability of cooperative behavior, several experiments have been performed  under very controlled conditions in the lab using artificial environments.\footnote{Note that studies explicitly considering the spatial extension of bacterial populations are not considered here but in Section~\ref{sec:spatially_extended_systems}.} 
Structures studied range from nanoscopic landscapes on a chip~\cite{Keymer:2006, Keymer:2008, Lambert:2014} to simple rearranging group structures~\cite{Buckling:2007, Chuang:2009, Chuang:2010, Becker_Wienand:2018, Kong:2018}.
The latter approach is especially useful for studying the influence of reoccurring population bottlenecks. 
For instance, these bottlenecks can account for species showing a life cycle. 
Such synthetic biology experiments, which help to clarify the mechanisms promoting cooperation in simple setups, may eventually  lead to a broader understanding of the cooperative behaviour in complex biological environments. 
In this review, we consider and analyze specifically the stability of pyoverdine production with bacteria growing in different well-mixed sub-populations. 
But before, we introduce the mathematical concepts required for the analysis of this case and cooperation within structured microbial populations in general.

\goodbreak

\section{Mathematical formulation of evolutionary dynamics and population growth}
\label{sec:mathematical_formulation}

In this section we introduce basic mathematical approaches to describe the evolutionary dynamics within populations. 
In particular, we will discuss evolutionary dynamics in well-mixed populations, considering frequency-dependent selection, demographic noise and population growth. 
Approaches to specifically analyze the evolution and maintenance of cooperation in structured populations build on these approaches and are discussed in the following Section~\ref{sec:evolinstructuredkinandgroup}.  
We tried to keep the mathematical formulation simple and focus on the underlying concepts instead. 
Readers with a focused interest on the biological aspects of cooperation can skip these sections and continue by reading about the specifics of pyoverdine production in Section~\ref{sec:pyoverdine}.

\subsection{Price equation}
\label{sec:price_equation}

\begin{figure}[t!]
\centering
\includegraphics[width=\textwidth]{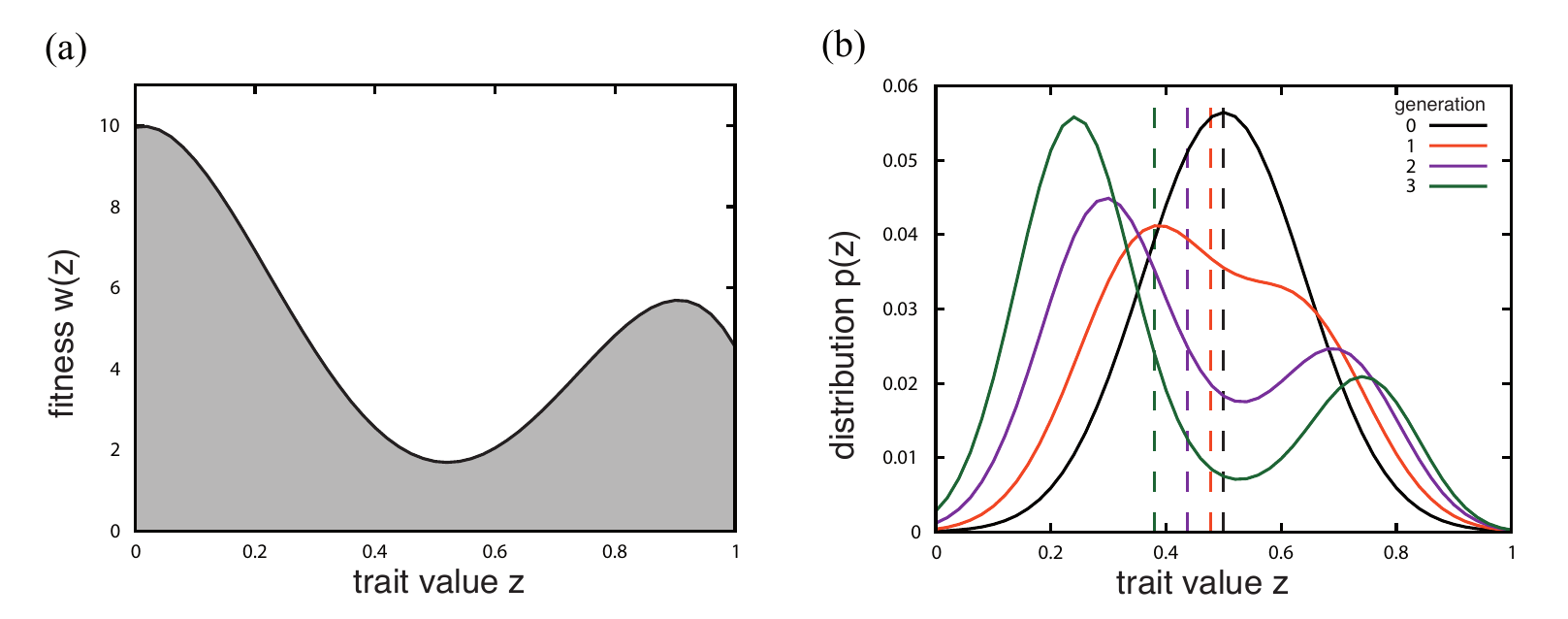}
\caption
[Evolutionary dynamics in a fitness-landscape. A population `climb up' the fitness landscape towards fitter states.]
{Evolutionary dynamics on a fitness-landscape. Different individuals in the population have a certain trait, characterized by the value $z$. By definition, the population evolves according to a trait dependent fitness factor $w(z)$. For illustration of the dynamics, consider a specific fitness function w(z) with two fitness peaks as shown in (a). Starting with a certain distribution of trait values within the population, black line in (b), the population evolves successively towards higher fitness values, colored lines in (b). More and more individuals adopt the fitter values over time, which for this specific example are given by a smaller and a larger $z$ value. The Price equation, Eq.~\eqref{eq:Price}, describes only the average value of the trait within the population, $\langle z \rangle$. This is shown by the dashed lines. This equation does not make any predictions about the distribution of trait values in the population, nor does it make any statements about the cause of underlying fitness-values. See text for further discussion.
\label{evol_fig:simple}}
\end{figure}

In $1970$, George Price proposed a generic equation to describe evolutionary dynamics including mutation and selection~\cite{Price:1970, Okasha, Hamilton:1975, Frank:1997, Chuang:2010, Zwietering:1990, Kerr:2009cf}. 
This equation is often used in the context of cooperation and we thus briefly introduce its motivation, following its simplest form. 
Let us consider a population containing $N$ individuals with the individuals labeled by an index $i\in\{1,2,...N\}$. 
Each individual is characterized by a trait, for example the body height or the weight, and a certain value of this trait  $z_i$ is assigned to each individual. 
At a given time point $t$, each individual has the abundance $h_i=1/N$ in the population.\footnote{The index $i$ can also be chosen to label the traits instead of the individuals. In this equivalent notation the abundance $h_i$  corresponds to the probability to find the trait $z_i$ in the population.} 
The average value of the trait is therefore given by\footnote{We choose the notation $\langle z_i\rangle$ instead of $\langle z \rangle_i$. This notation has some advantages when discussing evolution in structure populations, see Appendix~\ref{app:hamiltonsrule}.} $\langle z_i\rangle:=\sum_i h_i z_i=1/N\sum_i z_i$. 
Let us now consider how the average value of the trait changes over time. For a given time interval $\Delta t$ we describe the change by:
\begin{equation}
\langle\Delta z_i\rangle=\langle z'_i \rangle-\langle z_i\rangle.
\label{eq:der_P}
\end{equation}
Here,  $\langle z'_i\rangle$ is the average trait value at time $t'=t+\Delta t$. Again this average value can be calculated by $\langle z'_i\rangle=\sum_i h_i' z'_i$, but both the values of the trait, $z'_i$, as well as the abundances, $h_i'$, might have changed during the time interval. Let us further write the new trait of a certain individual $i$ as  $z_i'=z_i+\Delta z_i$. The new abundance follows is a consequence of different processes affecting replication and survival of individuals and their traits. To consider these differences, one can introduce fitness factors $w_i$\footnote{Importantly, these fitness factors are not the result of any specific theory considering reproduction or survival, they are simply introduced to consider differences in those processes and do not make any statement about how the differences emerge.}.\footnote{For example, consider reproduction within a certain time-interval: $w_i$ might be quantified by $2$ for a trait (or genotype) which allows reproduction once during the time interval, and by $0$ for a trait which dies during the time interval.} The new abundance of an individual $i$ can then be written by dividing its fitness factor, $w_i$, which corresponds to the number of individuals of type $i$ at time $t+\Delta t$, by the new population size, $N'$. The new population size can be calculated by multiplying $N$ with the average growth factor in the population, $\langle w_i\rangle=\sum_i w_i/N$. Taken together, the abundance at time $t+\Delta t$ is  given by  $h'_i=\frac{w_i}{N'}=\frac{w_i}{\langle w_i \rangle N}=\frac{w_i}{\langle w_i\rangle}h_i$. To derive the Price equation, we look at the average change of the trait value, Eq.~\eqref{eq:der_P}, which simplifies to:
\begin{eqnarray}
\langle\Delta z_i\rangle \langle w_i\rangle=&\langle{z_iw_i}\rangle-\langle z_i\rangle\langle w_i\rangle  +\langle\Delta z_i w_i\rangle  \nonumber\\
=& \text{Cov}[z_iw_i]+\langle \Delta z_i w_i\rangle.\label{eq:Price}
\end{eqnarray}
This is the Price equation. It is stating that traits which are positively correlated to the fitness factors increase in abundance, while others decline. As such, the Price equation makes a statement about the change of the population, provided fitness values are set; an illustration for evolution along a fitness landscape with continuous trait values is shown in Fig.~\ref{evol_fig:simple}.
However, it is important to realize that the Price equation does not provide any answers to what sets fitness. In this sense, the Price equation shares the same limitations as the famous phrase \emph{`survival of the fittest'}: During evolution the fittest individuals prevail, but the crucial question of how fitness values are set in a certain ecological setting is not considered.\footnote{Interestingly, Herbert Spencer was coining this phrase having a more specific meaning in mind, stressing that survival is also an important part of fitness.}

To be able to predictively describe evolution, one has to go beyond the Price equation and has to investigate evolutionary dynamics and establish fitness functions for the specific situation one studies. This requires the detailed consideration of the biological and ecological factors at play. Specific models are required to consider the different factors and their interplay. For  cooperation within microbial populations this requires for example the aspects of population growth, population structure, and the regulation of public goods. In general, many molecular and ecological features might be of relevance, making a full understanding of the dynamics of bacterial populations a multi-faceted and difficult endeavour.

Nevertheless, Price's equation with its consideration of evolutionary changes for given fitness values is an important starting point to think about evolution. We further discuss this aspect with the related approach of replicator dynamics in the next Section~\ref{sec:replicator_equation}. 
Moreover, the generic form of the Price equation also allows the consideration of evolution in structured populations and can thus also be used to think about the dilemma of cooperation under such conditions, see Section~\ref{sec:twolevelselectionandhamiltonsrule}.

\subsection{Replicator dynamics}
\label{sec:replicator_equation}

To describe the evolutionary dynamics over time, the replicator dynamics is also used often~\cite{Maynard, Hofbauer, Taylor:1978}. This approach considers the evolution of different species with different fitness values over time. It can be mathematically mapped to the Price equation as we further elaborate on in Appendix \ref{app:price_replicator}. Thus, the replicator dynamics does not provide any new concepts, but the involved replicator equations are often easier to read than Price equations. This is particularly true for the frequency-dependent situations discussed in the following Section~\ref{sec:evolutionary_game_theory}.

The replicator dynamics considers a population with $d$ different species (with distinct traits) where the relative abundance (frequency) of a species $k$ is given by $x_k$. Species differ in their fitness which are assumed to have fixed values $f_k$. The reasoning to set up a dynamic equation is then as follows: One expects that a given trait $k$ will increase in frequency if its fitness $f_k$ is larger than the average fitness in the population, $\langle f \rangle = \sum_k x_k \, f_k$. Conversely, $x_k$ will decrease if its fitness lies below $\langle f \rangle$. Thus, the evolutionary dynamics is often described by considering the following differential equation:
\begin{equation}
	\frac{d x_k}{dt}
	=
	f_k - \langle f \rangle 
	\, .
	\label{eq:replicator}
\end{equation}
This equation is known as the replicator equation, and adjusted replicator equation, respectively. $\frac{d x_k}{d t}$ denotes the change of $x_k$ over time. $x_k$ increases if the fitness of $k$ is larger than the average fitness, otherwise it decreases. Often, a slightly adjusted form of the replicator equation is also used:
\begin{equation}
	\frac{d x_k}{dt}
	=
	\frac{f_k - \langle f \rangle}
	     {\langle f \rangle}
	\, .
	\label{eq:replicator_adjusted}
\end{equation}
This adjusted replicator equation includes the same logic of selection and merely differs in the choice of the time-scale. As the Price equation, the replicator equations are heuristic and as such they do not provide arguments for choosing fitness values. Hence, they should be read as conceptual mathematical equations that capture selection \emph{given} certain fitness values.

\subsection{Frequency dependent fitness and evolutionary game theory}
\label{sec:evolutionary_game_theory}

The Price and replicator equations describe evolutionary dynamics for given fitness functions. As mentioned before, the exact fitness functions are typically specific to the biological scenario one is studying. However, some aspects of fitness functions can also be studied in more general frameworks. One particular aspect is the idea of frequency-dependent selection accounting for the possibility that the fitness of a certain trait may depend on the composition of the population~\cite{Levin:1988}. 

A quite successful approach to study such a scenario is evolutionary game theory (EGT). It was introduced in 1973 by Price and Maynard-Smith~\cite{Maynard, MaynardSmith:1973}, building on ideas and concepts developed in (classical) game theory by von Neumann~\cite{Neumann:1944}. 
In game theory, the success of certain strategies depends on the other participant participating in a game. 
Applied to evolution this means that the fitness of a given species depends on the composition of the whole population ~\cite{Maynard, Hofbauer, Taylor:1978, Nowakbook}. 

Consider a population consisting of individuals with $d$ different traits. Its composition is described by the vector $\textbf{x}= (x_1,x_2,...,x_d)$ where $x_k$ is the fraction of trait $k$. 
As discussed in Section \ref{sec:intro_game_theory}, one may represent the effect of the interactions between traits as \textit{payoffs}. 
These payoffs are summarised in the payoff matrix, ${\bf P}$, where the entries $P_{kl}$ characterise the gain of an individual with trait $k$ when interacting with an individual with trait $l$. 
Fitness of a trait, $f_k$, is then defined as a background (base) fitness (set to $1$) plus the average payoff obtained from interactions with all other individuals in the population
\begin{equation}
	f_\alpha
	=
	1 + s \sum_l P_{k l} \, 
	x_l
	\, .
\label{eq:fitness}
\end{equation}
Here, $s$ is the selection strength weighting the relative importance of the frequency-dependent fitness as compared to the base fitness of individuals.~\footnote{In principle the selection strength $s$ can be submerged into the payoff parameters, $P_{kl}$,but it is often introduced to be able to switch between neutral dynamics and selection, see the consideration of stochastic effects below.}

\paragraph{Public good game} Let us illustrate this for the two-trait public good game introduced in Section \ref{sec:intro_game_theory}. 
From the payoff matrix, we read off
\begin{align}
	f_C &= 1+s\,(b\,x-c) \, ,\\
	f_D &= 1+s\,b\,x \, .
\end{align}
Thus, the following replicator equation describes the dynamics,
\begin{equation}
	\frac{d x_k}{dt} x=-s \, c \, x\, (1-x)
	\, ,
\end{equation}
where $x$ is the fraction of cooperators. This equation becomes identical to the one stated in Section \ref{sec:intro_dilemma_cooperation} upon setting $s=1$ (which simply amounts to choosing a time scale).
Thus, the same conclusion holds: Independent of initial conditions, $\frac{d x_k}{dt}$ is always negative and cooperative individuals always die out.

These considerations can be generalised to all two-player games~\cite{Taylor:2004, NowakSigmund, Nowak:2004, Doebeli:2005, Hauert:2005, Antal:2006, Antal_Nowak:2008, Santos:2008, Cremer:2008, Cremer:2009, Frey:2010, Melbinger:2010, Szabo} as well as to games with more than two players, like three cyclically competing species; see e.g.\/ Refs.~\cite{Hofbauer, May_Leonard:1975, Frean:2001, Kerr_Riley:2002, Traulsen:2005, Reichenbach:2006, Claussen:2008, Cremer:2008, Berr:2009, Andrae:2010, Dobrinevski:2012, Parker_Kamenev:2009, Galla:2011, Knebel:2013, Knebel:2015} and references cited therein.

\subsection{The role of fluctuations}
\label{sec:fluctuations}

The approaches to describe evolutionary dynamics introduced above are deterministic and neglect the effect of randomness and noise. In reality, however, there are many sources of noise~\cite{Melbinger_Vergassola:2015}.  
For instance, the environment may not be constant but resources and other factors affecting the growth and death rates of individuals may fluctuate in time.  
This is sometimes referred to as \textit{extrinsic noise}.
Another source or noise which is rather \textit{intrinsic} is due to the fact that processes like reproduction, death, and mutation are random. 
This form of noise is also referred to as \textit{demographic noise}.  
In the following, we briefly consider the most important concepts which are needed in the following chapters; for a more detailed discussion of the different noise forms and the concepts to investigate their effect on evolutionary dynamics please refer to Ref.~\cite{Melbinger_Vergassola:2015}. 
To account for  random fluctuations in the size and the composition of a population, deterministic models in form of ordinary differential equation like the replicator dynamics are not sufficient. 
Instead, one needs to consider individual-based, stochastic models. 
These models are best formulated in terms of master equations; the interested reader may want to consult one of the standard textbooks on stochastic processes~\cite{Gardiner, Frey:2005}.
For simplicity, consider a population with only two different traits, $A$ and $B$, whose fitness is $f_A$ and $f_B$, respectively. 
\begin{figure}[!t]
\centering
\includegraphics[]{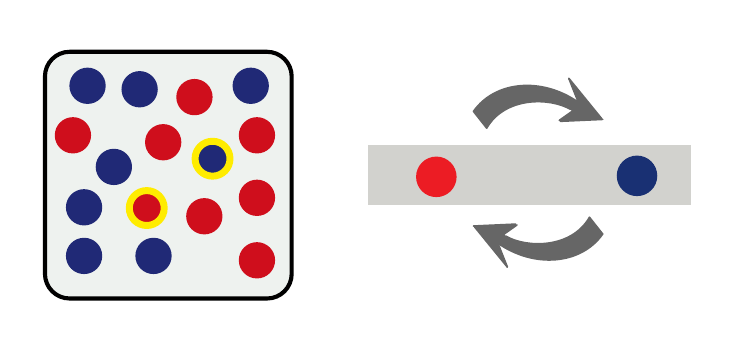}
\caption[The Moran process. Coupled birth and death events]
{
\textbf{Illustration of the Moran process as an urn model}. It describes the stochastic time evolution of well-mixed finite populations with constant population size. 
Here, as an example, we show a population consisting of two different traits (red and blue spheres). 
At each time step, two randomly selected individuals are chosen (left picture) and interact with each other.
Proportional to their fitness and abundance individuals are replacing each other as described in the main text (right picture).
\label{fig:moran}
}
\end{figure}
For a well-mixed population, the population may be envisioned as an urn containing $N$ individuals, $N_A$ belonging to A and $N_B$ belonging to B, see Figure~\ref{fig:moran}. 
The composition of the population is assumed to change stochastically according to some update rule, which mathematically is given in terms of a transition rates and which also considers the fitness differences within the population. A classical version is the Moran process~\cite{Moran} defined as follows: Two individuals are picked at random with probabilities given by their respective relative abundances. In the competition between these individuals, the fitness determines the likelihood of winning. The ensuing rates of replacement are given as
\begin{eqnarray}
	\Gamma_{B\to A}
	=
	\frac{f_A}{\langle f \rangle} \, 
	x_A \, x_B 
	\quad 
	\text{and}
	\hspace{.3cm} 
	\Gamma_{A\to B}
	=
	\frac{f_B}{\langle f \rangle}
	\, 
	x_A \, x_B 
	\, .
\label{eq:Moranrates}
\end{eqnarray}
For a more general overview on stochastic models in this field see e.g. Ref.~\cite{Blythe:2007}.

The full stochastic dynamics can be described in terms of a master equation for the probability distribution function $P(N_A,t)$:
\begin{equation}
	\frac{d P(N_A,t)}{dt}
	=
	\sum_S 
	\left[
	(\mathbb{E}_A^--1) \,
	\Gamma_{S\rightarrow A}
	+
	(\mathbb{E}_A^+-1) \,
	\Gamma_{A\rightarrow S}
	\right]
	P(N_A,t),
\label{eq:master}
\end{equation}
where $\mathbb{E}_A^\pm$ are step operators increasing/decreasing the number of individuals of trait $S$ by one~\cite{VanKampen:2001}, i.e.\/ $\mathbb{E}_A^\pm P(N_A) = P(N_A \,{\pm}\, 1)$.\footnote{For readers unfamiliar with this notation: the only important concept to get is that this equation describes the change in probability for individuals of type $A$ and $B$ to increase or decrease in abundance.} 
Solving such a master equation in closed form is almost never possible. 
Therefore, one either has to resort on numerical simulations or on approximation schemes.
The lowest order of such approximation, neglecting all correlations and fluctuations, corresponds to the set of equations studied in the previous section, which is often also referred to as a mean-field limit.
For a Moran process this mean-field approximation just leads to the adjusted replicator equation, Eq.~\eqref{eq:replicator}, not considering any fluctuations. 
To account for fluctuations, various approximations can be be employed\footnote{Most common approximations include the Kramers-Moyal expansion~\cite{Risken} or the Omega expansion proposed by van Kampen~\cite{VanKampen:2001} . While the first one works well for a constant population size, the second one is suitable for problems where this assumption is skipped. This is particularly the case for microbial populations for which sizes can strongly change, this is for example considered in Section~\ref{sec:random_drift_bact_pop}.}. A common one includes the derivation of the corresponding Fokker-Planck equation:
\begin{eqnarray}
	\partial_t P({\bf x},t)
	= 
	- \partial_x \alpha(x) P(x,t)
	+ \frac{1}{2}\partial_x^2 \beta(x) P(x,t)
\label{eq:FP}.
\end{eqnarray}
where
\begin{eqnarray}
	\alpha (x)
	=
	\frac{f_A x-f_B (1-x)}{\langle f \rangle}
	\, , 
	\quad \text{and} \hspace{.3cm} 
	\beta (x)
	=
	\frac{1}{N}
	\frac{f_A x +f_B (1+x)}
	     {\langle f \rangle}
	\, .
\label{eq:FPcoefficients}
\end{eqnarray}
Here, $\partial_t$ and $\partial_x$ describe derivatives in time and abundance, and the equation is now a partial differential equation. The first term describes \emph{directed drift}. In the large population size limit $N\to \infty$, this is the only remaining term and the dynamics is then given by the replicator equation, Eq.~\eqref{eq:replicator}: $\frac{d x}{dt} = \alpha(x)$. 
The second term describes the impact of demographic fluctuations. Confusingly, it is often referred to as \emph{random drift} in the literature. This term describes deviations from the deterministic solutions. The magnitude of this term scales as $1/N$. As an important consequence, fluctuations scale as $1/\sqrt N$.
Hence, the smaller a population is, the more pronounced is the role of fluctuations. In particular, if individuals occupy new habitats or undergo external catastrophes decimating their number, fluctuations gain importance and may alter the evolutionary outcome drastically. Another example for the importance of fluctuations are propagating fronts. At these front only a few individuals enter a new environment and fluctuations gain special importance~\cite{Brockmann2007, Hallatschek:2010, Hermsen:2010}. 

\subsection{Population growth}
\label{sec:popgrowth}

The formulations introduced up to now do not explicitly include varying population sizes. Strong growth is however an important characteristics of many microbial populations. In this paragraph we thus briefly introduce some models of population growth which we refer to later during this review.

One aspect particularly important for microbes is their fast and steady growth characteristics: Under nutrient rich growth conditions, cells can grow steadily over several generations~\cite{Neidhardt:1999}. In this case, increase in population size $N$ can be described well by exponential growth,
\begin{equation}
	\frac{dN}{dt}=\mu N \, ,
	\label{eq:exponential}
\end{equation}
with the growth rate $\mu$. This exponential growth eventually has to stop as the increasing nutrient consumption of the growing population will necessarily lead to a shortage in growth supplying nutrients. For microbes in well-defined culturing conditions within the laboratory, this behavior can be described well by Monod-kinetics and the explicit modeling of essential nutrient sources and their consumption by cells \cite{Monod:1949}. The major characteristics of this dynamics, fast growth when nutrient sources are highly abundant, and arrest of growth when nutrients run out, is already described well by the simpler logistic growth~\cite{verhulst_notice_1838}
\begin{equation}
	\frac{dN}{dt}=
	\mu \, N \, (1-N/K) \, .
	\label{eq:logistic}
\end{equation}
Here, the population size cannot exceed the carrying capacity $K$. In the following considerations, we use this simpler formulation to consider the evolutionary consequences of bacterial growth on the stability of cooperative traits.

\section{Evolution in structured populations - the frameworks of kin- and group-selection}
\label{sec:evolinstructuredkinandgroup}

The mathematical models described in the previous Section~\ref{sec:mathematical_formulation} describe evolutionary dynamics and population growth in well-mixed populations without any population structure.\footnote{To phrase it in game theory language: everyone can 'interact' with everyone else.} 
However, as we discussed in Section~\ref{sec:subseccooperationinthemicrobialworld}, microbial populations are highly structured and consist of many sub-populations. 
This structure might promote the clustering of cooperators with other cooperators and thus resolve the dilemma of cooperation; see Section~\ref{sec:factors_cooperationinmicrobialcooperation}. 
Historically, different consequences of population structure on evolution of cooperation have been described within the frameworks of kin-selection and group-selection.\footnote{We intentionally avoid to use the term `theory' here. While both frameworks are often called theories, this is in our opinion misleading giving the limited predictive power of both frameworks; see discussion below.} 
While we think that both frameworks do not provide particularly valuable insights into the requirements for stable cooperation, they are both commonly referred to, also in the literature on microbial populations. In this section, we thus introduce these frameworks, following our previous discussions~\cite{Melbinger:2011thesis,Cremer:2011thesis}. We further recommend the review by Damore and Gore~\cite{Damore_Gore:2012} on this topic. Though conceptually important, this section is not strictly required for the following discussions on cooperation in structured microbial cooperation; readers more interested in the specifics of microbial cooperation can continue reading from Section~\ref{sec:pyoverdine}.

Both the frameworks of kin-selection and group-selection, as well as their more evolved descendants (inclusive fitness and the framework of multi-level selection), have in common the attempt to provide general explanations for the stability of cooperative behavior. The different frameworks and the controversial debate about their relations have a long history. 
To better understand the controversy we first start with the historical context of both theories, before briefly introducing the mathematical approaches and their relations. We finish this section by commenting on the implications of the results, and why we think that these general conceptual frameworks provide no mechanistic insight about the stability of cooperative behavior.

\subsection{Group-selection}
\label{sec:group_selection}
The idea that selection might not only take place between individuals but also between larger entities or groups has been present in evolutionary theory right since its original formulation. For example, Darwin already proposed to consider such scenarios in his book `The Descent of Man'~\cite{Darwin:1871}. The idea is that the dilemma of cooperation can be overcome when groups with more cooperators have a selection advantage compared to groups with less cooperators. Due to this advantage on certain groups, the likelihood that a cooperator lives in a group with a relatively high fraction of cooperators is increased and therefore exploitation due to free-riders is reduced. Depending on the strength of this advantage and costs of cooperation, cooperative behavior might prevail.

A first mathematical model investigating group-selection was introduced by Sewall Wright~\cite{Wright:1931,Wright}. Since then, the concept of group-selection has been a controversial issue, mainly because of a careless assumption often made at the beginning: In many formulations, only evolutionary competition between groups was considered, favoring cooperation. The evolutionary dynamics within groups in which cooperators have a disadvantage compared to free-riders was mostly neglected, thereby creating biased results. Despite that, the idea of group-selection was widely used uncritically and especially put forward by Wynne-Edwards~\cite{Wynne-Edwards:1962} in the mid 20th century. In that period, the dilemma of cooperation was believed to be solved by selection on a group or even species level and the latter was even referred to as \emph{species selection}. However, this alleged success did not hold for long. Drastic criticism was formulated for example by George C. Williams~\cite{Williams:1966} and the idea of group-selection became less and less popular.
This view was shared by Maynard-Smith, who tried to corroborate this discussion using a mathematical model that considers two levels of selection (inter-group and intra-group) into account, the haystack model~\cite{MaynardSmith:1964}. He and others doubted that group-selection could be a general tool to explain cooperative behavior due to the restrictive conditions, such as strictly separated groups or a well-defined regrouping step, which were assumed~\cite{Wade:1978}. Since then, the dominant line of thinking has been: group-selection is possible in principle but practically not relevant. If the term group-selection is used in the strict sense of Wynne-Edwards, then this statement is certainly true and already indicates for the first time the semantic confusions which drive the debate. 
However, if one considers group selection as a term stressing the existence of sub-populations like groups and selection acting also on these entities, a benefit for cooperating individuals cannot be controversial, see e.g. Refs.~\cite{Wilson:1977, Wilson:1977a, Wade:1978, Wilson:1987, Silva:1999, Fletcher:2004, Jansen_Minus:2006, Killingback:2006, Traulsen:2006a, Fletcher:2007, Ichinose:2008, Traulsen:2008}.
 
Nowadays, many extensions of the idea of group-selection, including a weaker definition of the term `group', have been proposed to explain cooperative behavior, often including selection on many levels and the framework of \emph{multi-level selection}, see e.g.\/ Refs.~\cite{DSWilsongroup, Wilson:1977, Wilson:1983, Wilson:1987, Sober:1998, Silva:1999, Kerr:2002, Fletcher:2004, Killingback:2006, Traulsen:2006a, Okasha}. Still, they encounter much criticism, especially from the proponents of kin-selection. In Section~\ref{sec:twolevelselectionandhamiltonsrule} we will briefly discuss a particular situation of this framework in which individuals are arranged in groups and then comment briefly on the debate. 
But first, we consider the historical context of kin-selection.

\subsection{Kin-selection}
\label{kin_selection}

In contrast to group-selection, the framework of kin-selection emphasizes relatedness and promotes Hamilton's rule to explain cooperative behavior.
The idea is that interactions mostly among related individuals (same kin), leads to a clustering of cooperating individuals and decreases the probability of being exploited by non-related, non-cooperating individuals. 
Historically, kin-selection was motivated by the notion that one cannot only spread one's own genes by
reproduction but also by supporting relatives sharing one's genetic material. This line of thinking was already acknowledged by Fisher and Haldane~\cite{Fisher,Haldane}.

Haldane in particular is often cited for providing the idea of kin-selection in its simplest form, suggesting that kinship alone is sufficient to explain cooperative behavior. Further, the idea of kin-selection  is often summarized with a trivialized version  of 'Hamilton's rule', stating that cooperation is beneficial if the benefit $b$ of a cooperative behavior weighted by the genetic relatedness $r$ (overlap of genes between interacting individuals) exceeds the cost $c$ for providing the cooperative behavior:
\begin{equation}
    b\cdot r>c
    \label{eq:hamiltonsrule_simple}
\end{equation}
However, the involved quantities are not simply constants but complex non-linear functions depending on several continuously changing variables. Oversimplified interpretations of the inequality, as often discussed in textbooks, lead to enormous misinterpretations concerning the reasons of cooperation.  

Earlier population geneticists thinking about kin-selection were already aware of these subtle issues. For example, they rarely talked about relation in the strict kinship sense (like related individuals belong to the same family) and were fully aware that more specific considerations are needed to make statements about the stability of cooperative behavior. For example, Haldane himself humorously wrote in his article ``Population Genetics'' published in 1955~\cite{Haldane:1955}: 
\begin{quote}
 \textit{``Let us suppose that you carry a rare gene that affects your behavior so that you jump into a flooded river and save a child, but you have one chance in ten of being drowned, while I do not possess the gene, and stand on the bank and watch the child drown. If the child's your own child or your brother or sister, there is an even chance that this child will also have this gene, so five genes will be saved in children for one lost in an adult. If you save a grandchild or a nephew, the advantage is only two and a half to one. If you only save a first cousin, the effect is very slight. If you try to save your first cousin once removed the population is more likely to lose this valuable gene than to gain it. (...) It is clear that genes making for conduct of this kind would only have a chance of spreading in rather small populations when most of the children were fairly near relatives of the man who risked his life.''}   
\end{quote}

Further, Hamilton's own motivation for the inequality now bearing his name clearly illustrates that the quantities involved in this relation are not simply constant parameters but very complex quantities; see also discussion below, Section~\ref{sec:note_hamiltonsrul}.
As stressed by Haldane, Hamilton, and others, additional ecological or biological conditions are needed to ensure that related individuals mostly interact with each other, and that such a preferential interaction is maintained over time. 
Most prominently, Hamilton summarized some of the necessary conditions~\cite{Hamilton:1964,Hamilton:1964}. 
He specifically mentioned the ability of individuals to recognize their own kin, or the environmental condition leading to 'viscous populations' and by such to an increased likelihood of related individuals interacting with each other. 
While the ability to recognize related cooperating individuals might be hard to maintain (but see discussion Section~\ref{sec:factors_cooperationinmicrobialcooperation}), many ecological conditions might lead to the clustering of more related microbes.  
As such, the framework of kin-selection shares similar ideas to that of group-selection. 
Indeed, the inclusive fitness framework, a more general mathematical consideration of kin-selection dating back to Hamilton, resembles the mathematical formulation of multi-level selection, a general mathematical description of group-selection. In the following we discuss this relation, by considering  evolution in a simply structured population with individuals belonging to different groups first.

\subsection{Two levels of selection and derivation of Hamilton's rule}
\label{sec:twolevelselectionandhamiltonsrule}

\begin{figure}[!t]
\begin{center} 	
\includegraphics[width=0.9\linewidth]{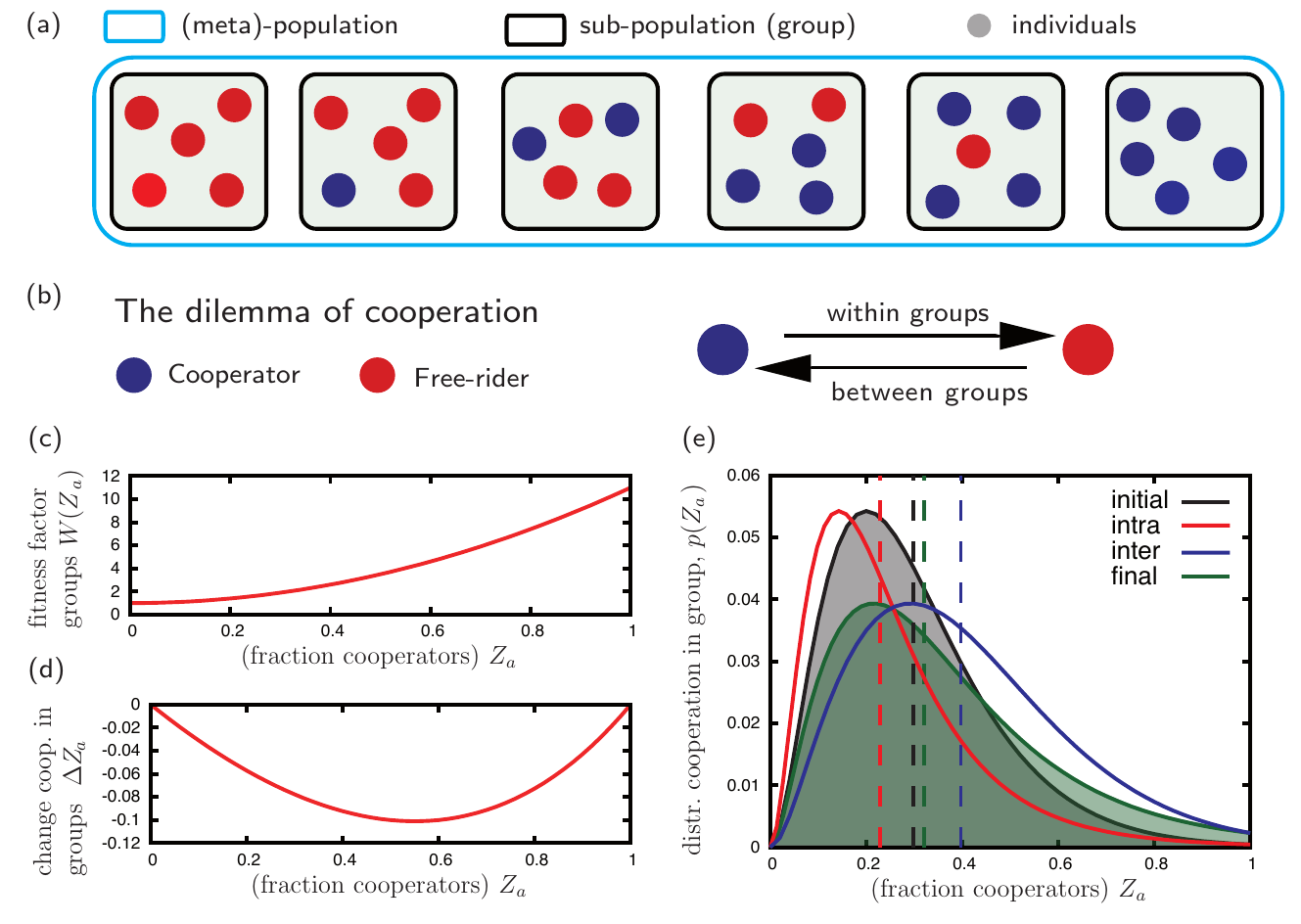}	
\end{center}
\caption[Selection in a two level setup. The interplay of intra- and inter-group evolution determines the total evolutionary outcome.]
{
\textbf{Selection on two-levels in group-structured populations}. (a) A population with individuals belonging to different groups (sub-populations). (b) For the dilemma of cooperation, selection within the sub-populations selects for free-riders while competition between groups selects for more cooperative groups. In the multi-level view of selection, intra-group evolution selects for free-riders, while inter-group evolution selects for cooperative behavior. (c-e). The outcome depends on the exact comparison of both selection processes. Here one example is shown, following the two-level Price equation approach (details in Appendix~\ref{app:hamiltonsrule}). (c) and (d), exemplary fitness on the group-level and the evolutionary dynamics within each group depend on the fraction of cooperators within the groups $(Z_\alpha)$. (e) Consider an initial distribution of the fraction of cooperators within groups (black line, average shown as dashed line). This distribution changes according to the selection dynamics within groups and the competition between groups. If only selection between groups is considered, then one obtains the new distribution shown in blue in (e): cooperation increases within each group and on average.  If only selection within groups is considered, the new distribution is shown red in (e): cooperation decreases within each group and on average. The total outcome considering both levels is shown in green: while cooperation decreases within each group, the average level of cooperation still increases. Thus, population structure can promote cooperation. However, this example also illustrates that selection within groups leads to a lower level of cooperation within each group. For cooperation to be stably maintained additional mechanisms are thus required to keep cooperation levels within groups high. Thus, \emph{to understand the evolutionary stability of cooperation, more specific considerations are required, going beyond the abstract considerations of the Price equation, multi-level selection or inclusive fitness}. Adapted from Ref.~\cite{Cremer:2011thesis}.
\label{fig:two_level}
}
\end{figure}

Historically, selection within structured populations has been analyzed by the Price equation. Here we illustrate the idea behind this approach by looking at a simple population structure with two types of individuals, $A$ and $B$ assigned to different sub-populations (groups); see Fig.~\ref{fig:two_level}(a) for an illustration. This scenario was first considered by Price and Hamilton~\cite{Price:1970, Hamilton:1975}; see also~\cite{Frank:1997, Frank:1998, Okasha, Chuang:2010} for detailed reviews. 

Within each group there is selection towards those individuals with a higher reproduction rate. At the same time,  groups may do better or worse, depending on their internal composition (and possibly the competition with other groups). Phrasing it differently, there is selection on two levels: within the group (intra-group level) and between groups (inter-group level). If selection on both levels favours one type of individual over the other, then the evolutionary outcome is obvious and the interplay between both levels only sets the time scale of selection. The situation is more interesting if the different levels favour different traits. Consider the dilemma of cooperation as a specific example were individuals are either public good providing cooperators (type $A$) or free-riders (type $B$). In this situation, selection within groups favours the free-riders which save the costs for providing the public good. At the same time, selection between groups can lead to an advantage of more cooperative groups, see ~\ref{fig:two_level}(b) for an illustration.  Whether type $A$ or type $B$ increases its global fraction in the population depends on the structuring of the population and the interplay of the two different levels. This situation can be mathematically analysed with the Price equation and an example is illustrated in Figure~\ref{fig:two_level}(c-e) for a specific assumption of fitness advantages within groups and the competition between groups. The illustrated scenario shows a scenario of Simpson's paradox~\cite{Chuang:2009,Chuang:2010}. While free-riders increase within each group. The average level of cooperation within the population increases. 

The Price equation approach to compare evolution on two levels is outlined in Appendix~\ref{app:hamiltonsrule}. Result of this analysis is a general form of Hamilton's rule, an inequality comparing the benefits arising from cooperation with its total costs and relation,~\cite{Price:1970, Hamilton}:
\begin{equation}
\mathcal{R}\mathcal{B}>\mathcal{C}
\label{eq:Hamiltonsrulegeneral}
\end{equation}
As with the simplified form stated in Eq.~\eqref{eq:hamiltonsrule_simple}: for cooperation levels to increase the benefit $(\mathcal{B})$ weighted by the relatedness $(\mathcal{R})$ has to exceed the costs $(\mathcal{C})$.

\subsection{A note on Hamilton's rule}
\label{sec:note_hamiltonsrul}

Importantly, Hamilton's rule even in its general form, Eq.~\ref{eq:Hamiltonsrulegeneral}, is prone to oversimplified, highly misleading interpretations. First, note that all three quantities in Eq.~\eqref{eq:Hamiltonsrulegeneral} are complex functions which depend on the current state of the system (definitions of the functions are provided in Appendix \ref{app:hamiltonsrule}). In particular, these functions are not directly measurable quantities with constant values for costs, benefits, and relatedness. Instead, they change over time and depend on the state of the population. In particular, the variance terms defining relatedness $\mathcal{R}$ rely on the details of the underlying population structure. 
Simpler versions of Hamilton's rule are only obtained for very specific cases. For example, as pointed out by Chuang et al.~\cite{Chuang:2010}, $\mathcal{B}$, $\mathcal{R}$, and $\mathcal{C}$ can only be assumed to be constant numbers if the fitness terms $f_{i,m}$ and $G_{m}$ depend linearly on the frequencies $\lbrace x_m\rbrace$ ---a condition hardly fullfilled in microbial populations~\cite{Chuang:2010, Becker_Wienand:2018} (see Appendix~\ref{app:hamiltonsrule} for notation).

Overall, the existence of Hamilton's rule further supports the idea that the fitness disadvantage of cooperative behaviour can in principle be overcome by advantages on higher levels of selection (\emph{e.g.} the group level). However, it should not be misunderstood as a rule providing any mechanistic insights on how population structures ensures the stability of cooperation: While the condition can always be stated for a specific situation at a certain time, its predictive power for evolutionary outcomes is rather limited. This is not surprising, as the Price equation itself says nothing about the detailed dynamics but describes the change of expectation values during a fixed time interval, provided the fitness values for that certain time-window are given. It does not provide any insights into how the assortment of cooperators is maintained (see also the discussion in Section~\ref{sec:mathematical_formulation}). 
For a real understanding of the mechanisms promoting assortment cooperation in a specific situation, one has to specifically consider the different biological and ecological aspects at play.

Finally, given this context of Hamilton's rule, we also think that the ongoing debate between proponents of the group- and kin-selection frameworks is not very fruitful. This includes for example more recent debates on the role of inclusive fitness,~e.g.~\cite{NowakWilson,Boomsma:2011,Ferriere:2011,Strassmann:2011,Herre:2011,Abbot:2011,Nowak:2011}. Within their general formulations, multi-level selection and inclusive fitness calculations are mathematically equivalent (see discussion above and ~\cite{Lehmann:2007,Traulsen:2009}), and they attempt to describe evolution on the same, very generic level. Differences are merely in the emphasis on different terms (for example \emph{relatedness} or \emph{groups}). There are several recent reviews on the general state of the debate, \emph{e.g.}~\cite{West:2007,Wilson:2008,West:2007b,Wilson:2005,Foster:2006,Birch:2017kq}, so we just mention how much in our opinion both theories resembles each other. 

While  group-selection focuses on structure to explain cooperation, kin-selection focuses on relatedness as the reason for cooperative behaviour. But actually relatedness as well as structure are necessary for both, which therefore strongly resemble each other: On the one hand, groups can only favor cooperation if some of them have a higher level of cooperators and thereby a selection advantage compared to other groups. In terms of kin-selection this differences in the group composition correspond to a relatedness. 

On the other hand, the relatedness, as we have learned from Hamilton's rule, is not an absolute value (like a difference in the genome) but depends on the variance in the composition of all sub-populations. Thereby also structure is essential for this quantity. Crucially, both frame-works need additional mechanisms to ensure a stable high level of relatedness (or the existence of more cooperative groups). 
Without such a mechanisms, the difference in groups (or relatedness) declines over time as populations in sub-populations fixate and non-cooperators dominate. Thus, to understand these mechanisms, we have to leave the general formulations of kin- and group-selection, and study more specifically the evolutionary dynamics for the (microbial) populations we are interested in. What are the microscopic reasons leading to an (inclusive) fitness and group structures favouring cooperation? What are the dynamic processes underlying both frame-works? How can cooperative behaviour have emerged in the first place? These and many other questions still lack a satisfactory answer. 

\section{Pyoverdine - a public good in pseudomonas populations}
\label{sec:pyoverdine}

Prominent examples of cooperating microbes are siderophore-producing pseudomonads, with the best known group of siderophores being the fluorescent peptidic pyoverdines~\cite{Buckling:2007, Cornelis:2010, Cezard:2015, Ringel_Bruesner:2018}. 
Pyoverdines serve as iron-scavenging pigments and thereby enable cells to uptake iron, an essential element for almost every living organism, including pseudomonads~\cite{Cezard:2015}. We here discuss the biology of pyoverdines in more detail as we will use it as an example to illustrate how the  regulatory control of public good production and their chemical properties are important to consider. As explained in more detail in Section \ref{sec:siderophores}, the production of siderophores is a cooperative trait. 
It is beneficial for a population growing under iron-limiting conditions: wild-type pyoverdine producers grow faster and reach higher population densities than related non-producers, when strains are grown individually.
It is metabolically costly to the producer, and siderophores can be utilized by producers and non-producers~\cite{Griffin_West:2004, Kummerli_Gardner:2009, Kummerli_Griffin:2009}. 
As a consequence, a social dilemma arises since non-producers have a growth advantage compared to producers when grown in mixed culture.\footnote{As we noted already in Section \ref{sec:evolution_mircrobes}, there are several other examples for public goods in microbial systems, e.g.\/ sticky polymers connecting a microbial colony  as observed with \textit{P. fluorescens}~\cite{Rainey:2003} or invertases hydrolysing disaccharides into monosaccharides in the budding yeast \textit{Saccaromyces cerevisiae}~\cite{Greig:2004, Gore:2009, MacLean:2010}.}  

There are additional aspects of siderophore production which make it a versatile model system to study the influence of ecological and environmental factors on the maintenance and evolution of cooperation. 
First, the metabolic load put on the production of siderophores can be regulated by the amount of available iron~\cite{cornelis_iron_2009, Imperi:2009}. 
In fact, in \textit{P. aeruginosa} pyoverdine production decreases with increasing iron concentrations and ceases completely under high iron supplementation (FeCl$_3 \geq 50 \mu$M)~\cite{Kummerli_Jiricny:2009, Leinweber:2018}. 
Second, siderophore production is down-regulated in pyoverdine-rich environments~\cite{Kummerli_Brown:2010}. Third, pyoverdines are fluorescent and can therefore be measured to high accuracy. 
This property allows also easy discrimination of producers and non-producers after co-cultivation.  
Finally, fluorescent pseudomonads are of great importance in medicine (e.g. lung infections by \textit{Pseudomonas aeruginosa}~\cite{Parkins:2018} and for bioremediation (e.g.\/ \textit{Pseudomonas putida} colonizing plant roots)~\cite{Molina:2005}. 
Therefore, a better understanding of the role of population dynamics including competition and development of heterogeneity during colonisation of the respective environments is expected to support development of new strategies of medical therapy as well as of environmental protection.

In the following, we give a more detailed account of the regulatory network of pyoverdine synthesis and secretion and review recent progress in establishing \textit{Pseudomonas putida} as a model system to systematically study cooperation in bacterial populations. 
We will use this information in the following section to construct a mathematical model for populations containing producing and non-producing strains. There, we will examine how the biological details discussed in this section are important to take into account when considering the evolutionary dynamics of public good production.

\subsection{Biological and biochemical characterisation of pyoverdines} 
\label{sec:siderophores}

\begin{figure}[!t]
\centering
\includegraphics[width=0.9\textwidth]{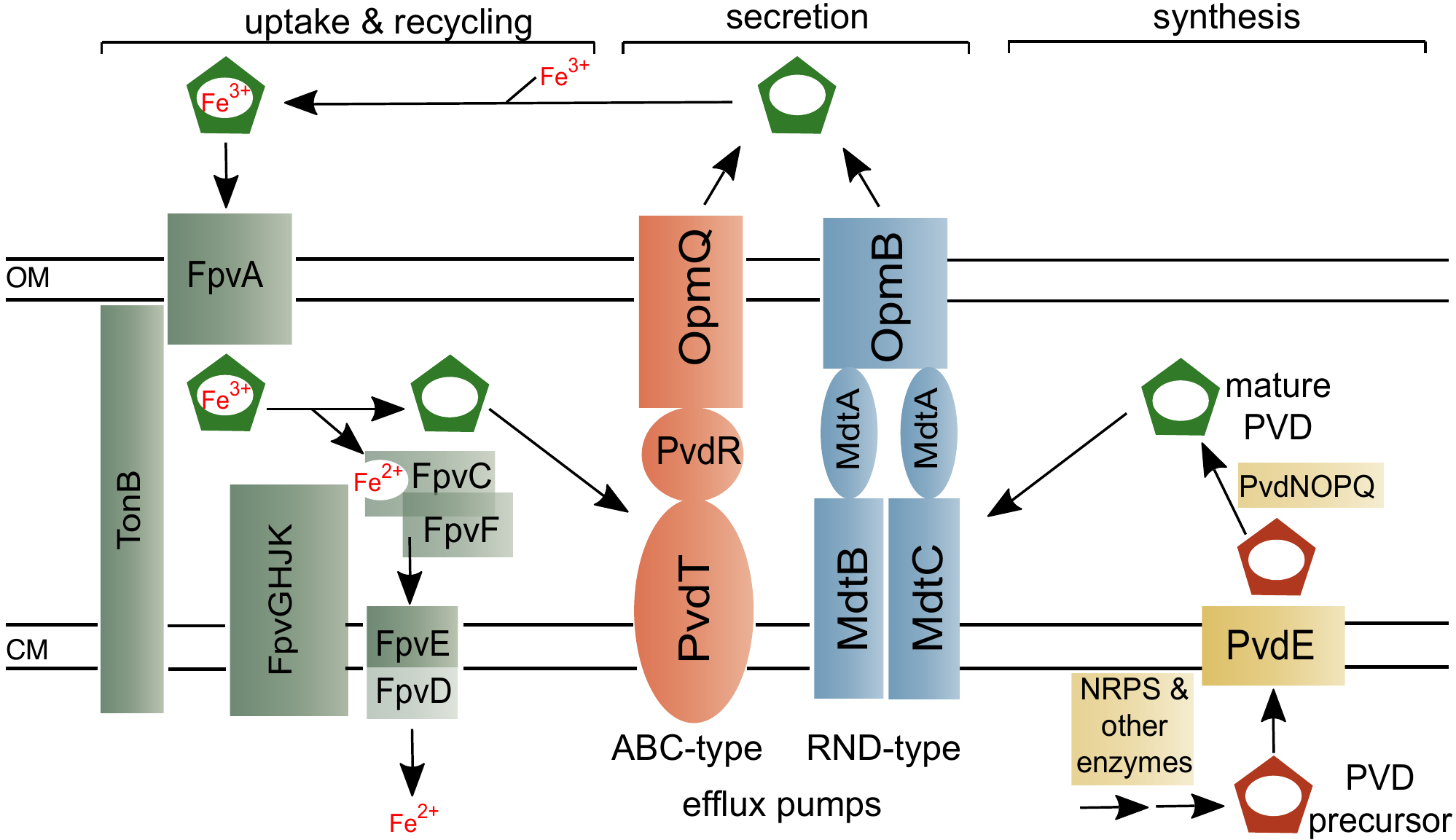}
\caption{
\textbf{Model of pyoverdine synthesis, secretion, uptake and recycling in pseudomonads}. 
Non-ribosomal peptide synthetases (NRPS) produce a peptide precursor (red pentagon) that is further modified and acylated at its N terminus~\cite{Ringel_Bruesner:2018}. 
All the enzymes seem to form a large complex termed siderosome that is associated with the inner leaflet of the cytoplasmic membrane (CM) preferentially at cell poles~\cite{Gasser:2015, Imperi:2013}. 
The resulting precursor is transported across the CM most likely by the ABC transporter PvdE~\cite{Yeterian:2009}. 
Various periplasmic enzymes (e.g., PvdN, PvdO, PvdP, PvdQ) participate in further modifications and chromophore formation to finally yield mature fluorescent pyoverdine~\cite{Ringel_Bruesner:2018}. 
Transport of newly synthesized pyoverdine from the periplasm across the outer membrane (OM) is suggested to involve PvdRT-OpmQ, a tripartite efflux pump of the ATP-binding-cassette (ABC)-type~\cite{Hannauer:2010}. 
Recently, also the resistance-nodulation-division (RND)-type efflux pumps MdtABC-OpmB has been implicated in the process~\cite{Henriquez:2019}. Secreted pyoverdine binds Fe$^{3+}$ in the environment, and the pyoverdine-Fe$^{3+}$ complex is transported back into the periplasm by the OM receptor FpvA in a TonB-dependent manner~\cite{Imperi:2009}. 
Fe$^{3+}$ is reduced to Fe$^{2+}$ by the FpvGHJK complex, released from pyoverdine and via the binding proteins FpvC/FpvF and the ABC transporter complex FpvD/FpvE transported into the cytoplasm~\cite{Ganne:2017, Brillet:2012}. 
The remaining pyoverdine is transported back to the extracellular space to capture more iron~\cite{Imperi:2009}.}
\label{fig:HJ1}
\end{figure}

Siderophores are organic compounds that are synthesised and secreted by bacteria into the environment to scavenge ferric iron when it becomes scarce. 
The resulting iron-siderophore complexes are taken up by the bacteria, and iron is incorporated into bacterial proteins~\cite{Braun_Hantke:2011, Cornelis:2010, Miethke:2007}. 
Iron is a cofactor of different enzymes of central metabolic pathways and involved mainly in redox reactions (e.g., of the respiratory chain). The described  strategy of iron acquisition is thus essential for bacterial growth and survival in many native habitats since ferric iron has a very low solubility and is bound to proteins in hosts~\cite{Braun_Hantke:2011, Cornelis:2010}. 
 For example, fluorescent pseudomonads produce a group of siderophores called pyoverdines (as mentioned before) that are crucial for colonization of other organisms including plants, animals, and humans by pathogenic and non-pathogenic \textit{Pseudomonas} strains ~\cite{Miethke:2007}. Besides iron acquisition, pyoverdines are involved in the resistance against heavy metals~\cite{Schalk:2011} and oxidative stress~\cite{Jin:2018}, and interactions of ferri-pyoverdine complexes with other redox active compounds can provide access to phosphates, trace metals, and organic compounds in minerals~\cite{Price-Whelan:2006, Zhang:2013}. 
Pyoverdines are composed of three different components: (i) a dihydroxychinoline chromophore that is responsible for the fluorescence of the molecule, (ii) an acyl side-chain bound to carbon 3 of the chromophore, and (iii) a peptide moiety (6-14 amino acids) that is strain specific~\cite{Cezard:2015, Ringel_Bruesner:2018}.
The production of the pyoverdine starts in the cytosol with the synthesis of a precursor (ferribactin) by non-ribosomal peptide synthases~\cite{Mossialos:2002}. 
After transport into the periplasm by an ABC transporter, the precursor is modified to yield mature fluorescent pyoverdine~\cite{Ringel_Bruesner:2018}.
The latter is secreted into the extracellular environment by tripartite efflux pumps belonging to the ABC and RND classes of transporters~\cite{Henriquez:2019, Hannauer:2010} (Fig.~\ref{fig:HJ1}).

\paragraph{The pyoverdine synthesis and secretion system} 
Synthesis and secretion of pyoverdine is estimated to require 26 high energy phosphates \cite{Sexton:2017}. 
Up to 15\% of the ATP necessary to synthesize all constituents of a bacterial cell are estimated to be invested in pyoverdine production~\cite{Sexton:2017}. 
Major cost saving strategies of pseudomonads involve recycling of pyoverdine as well as a tight regulation of pyoverdine production. 
Importantly, once synthesized, pyoverdine is not consumed but recycled. 
After uptake of the pyoverdine-Fe complex into the periplasm, ferric iron is reduced and released from the siderophore. 
To capture more iron, the latter is pumped back into the environment probably by the same efflux pumps that translocate newly synthesized pyoverdine~\cite{Hannauer:2010, Imperi:2009} (Fig. \ref{fig:HJ1}). 
The central element of the regulatory network is the ferric uptake regulator (Fur) protein, an intracellular iron sensor repressing the synthesis of transcriptional activators of the iron acquisition system in its iron bound form. 
Under conditions of iron limitation, Fur is released from target promoters leading to the expression of a set of sigma factors~\cite{cornelis_iron_2009}. 
PvdS is one of these sigma factors and functions as activator of pyoverdine synthesis gene expression~\cite{Chevalier:2018, Llamas:2014}. 
This function of PvdS is inhibited by the anti-sigma factor FpvR which is located in the inner membrane and interacts with the outer membrane receptor FpvA. 
Binding of external pyoverdine-Fe to the receptor alters the protein-protein interactions and leads to the release of PvdS and stimulation of the expression of pyoverdine synthesis genes~\cite{Chevalier:2018, Llamas:2014}. 
By this means, pseudomonads can steadily adapt expression of pyoverdine synthesis genes to changing environmental conditions. 

\begin{figure}[!t]
\centering
\includegraphics[width=0.7\textwidth]{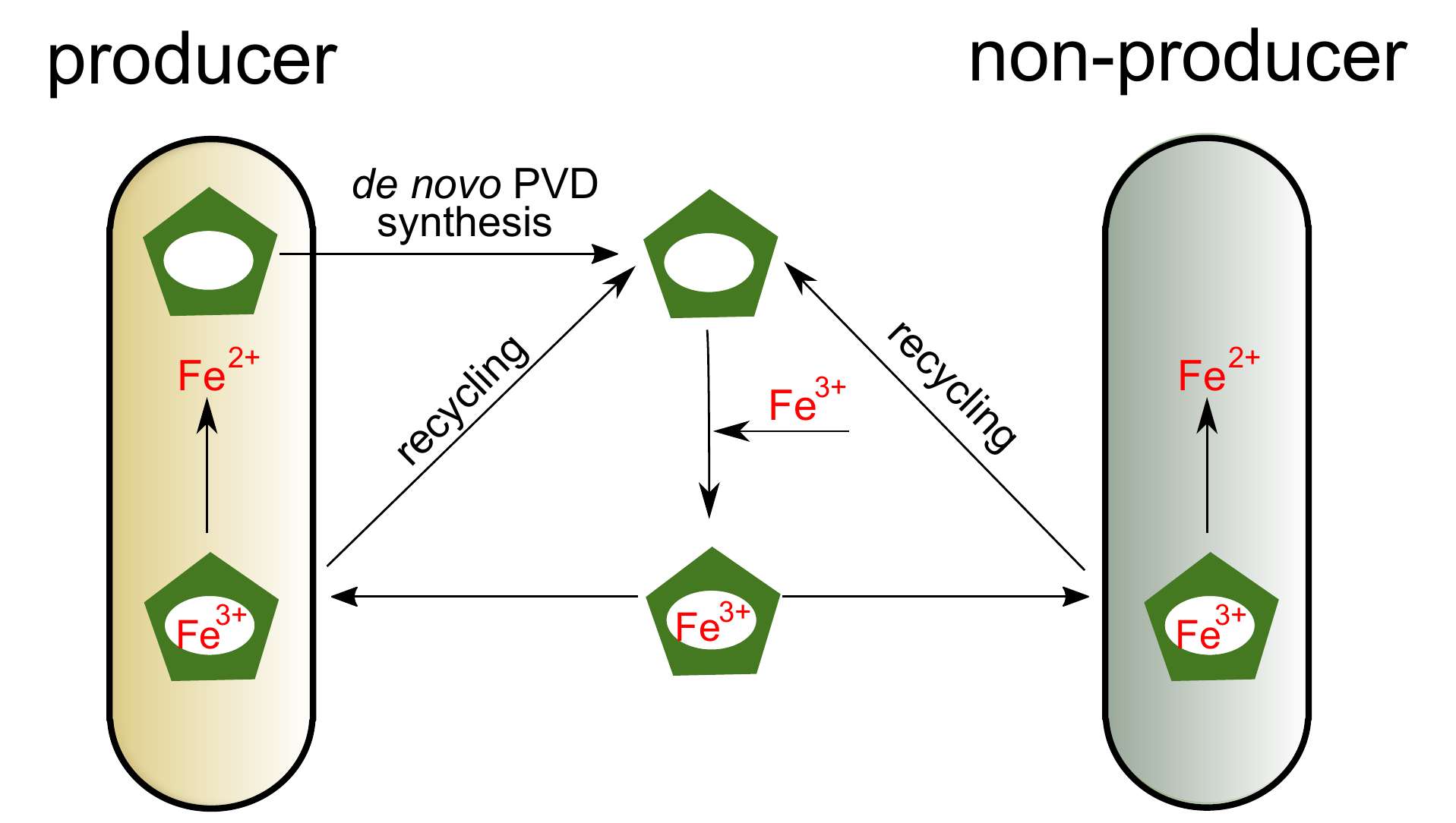}
\caption{
\textbf{Pyoverdine acts as a public good}. 
pyoverdine is produced by a cooperator (yellow) and secreted to the extracellular space where it can bind iron (Fe$^{3+}$). 
The complex is recognised and internalised by both producer and non-producer (green), even when the latter one does not pay the cost of its production (selection advantage). 
In the periplasm, iron is released and pyoverdine is transported back to the extracellular space to repeat the cycle (siderophore recycling).
}
\label{fig:HJ2}
\end{figure}

\paragraph{Pyoverdines represent public goods} 
When secreted into the environment, pyoverdines are expected to be used by producing (cooperators) and non-producing cells (free-riders)~\cite{West_Diggle:2007} (Fig.~\ref{fig:HJ2}). 
Indeed, it has been shown experimentally that secreted pyoverdine simultaneously supports growth of producers and non-producers under iron depletion (see e.g. Refs.~\cite{Kummerli:2015}). 
However, conditions inhibiting diffusion of pyoverdine such as spatial restrictions, high medium viscosity or other not well-mixed settings may prevent homogenous spreading over the entire population and support a (partially) private use of it~\cite{Dobay:2014}. \
For example, cell-cell contacts have been shown to confine public diffusion in \textit{P. aeruginosa} microcolonies~\cite{Julou:2013}. 
Furthermore, it has been shown that \textit{P. aeruginosa} regulates the secretion of iron-scavenging siderophores in the presence of environmental stresses, reserving this public good for private use in protection against reactive oxygen species when under stress~\cite{Jin:2018}. 
Consequently, if pyoverdine is considered a public good, detailed environmental conditions must be taken into account. In particular, costs and benefit of pyoverdine as a public good cannot simply be described with two parameters, as one typically do in evolutionary game theory (Section~\ref{sec:evolutionary_game_theory}).

\subsection{\textit{Pseudomonas putida} as an experimental  model system}
\label{sec:model_system_pseudomonas}

As shown recently \cite{Becker_Wienand:2018},  the soil bacterium \textit{P. putida} KT2440 is a versatile experimental model systems to investigate the social role of public goods.
The main advantage of this system is that it synthesises only a single type of siderophore~\cite{Matthijs:2009} mediating all cell-cell interactions, and does not produce any known quorum-sensing molecules that might otherwise interfere with the social interaction \cite{DosSantos:2004, Niewerth:2011, cornelis_iron_2009}. Wild-type \textit{P. putida} KT2440 controls pyoverdine production through a complex regulatory network, as explained in the previous section.
It allows cells to continually adapt their pyoverdine production to the availability of iron \cite{Matthijs:2009, Swingle:2008}. 
From the perspective of designing a well-controlled experimental model system, this regulation represents a downside, as it obscures the costs of pyoverdine production by affecting other processes. 
One way to circumvent this is to generate strains that constitutively produce pyoverdine (e.g., \textit{P. putida} KP1) \cite{Becker_Wienand:2018} and study populations where this strain competes against a non-producer strain carrying, e.g., an inactivated non-ribosomal peptide synthetase gene that inhibits pyoverdine synthesis and is otherwise isogenic (e.g., \textit{P. putida} 3E2) \cite{Matthijs:2009}.

\begin{figure}[!t]
\center\includegraphics[width=\linewidth]{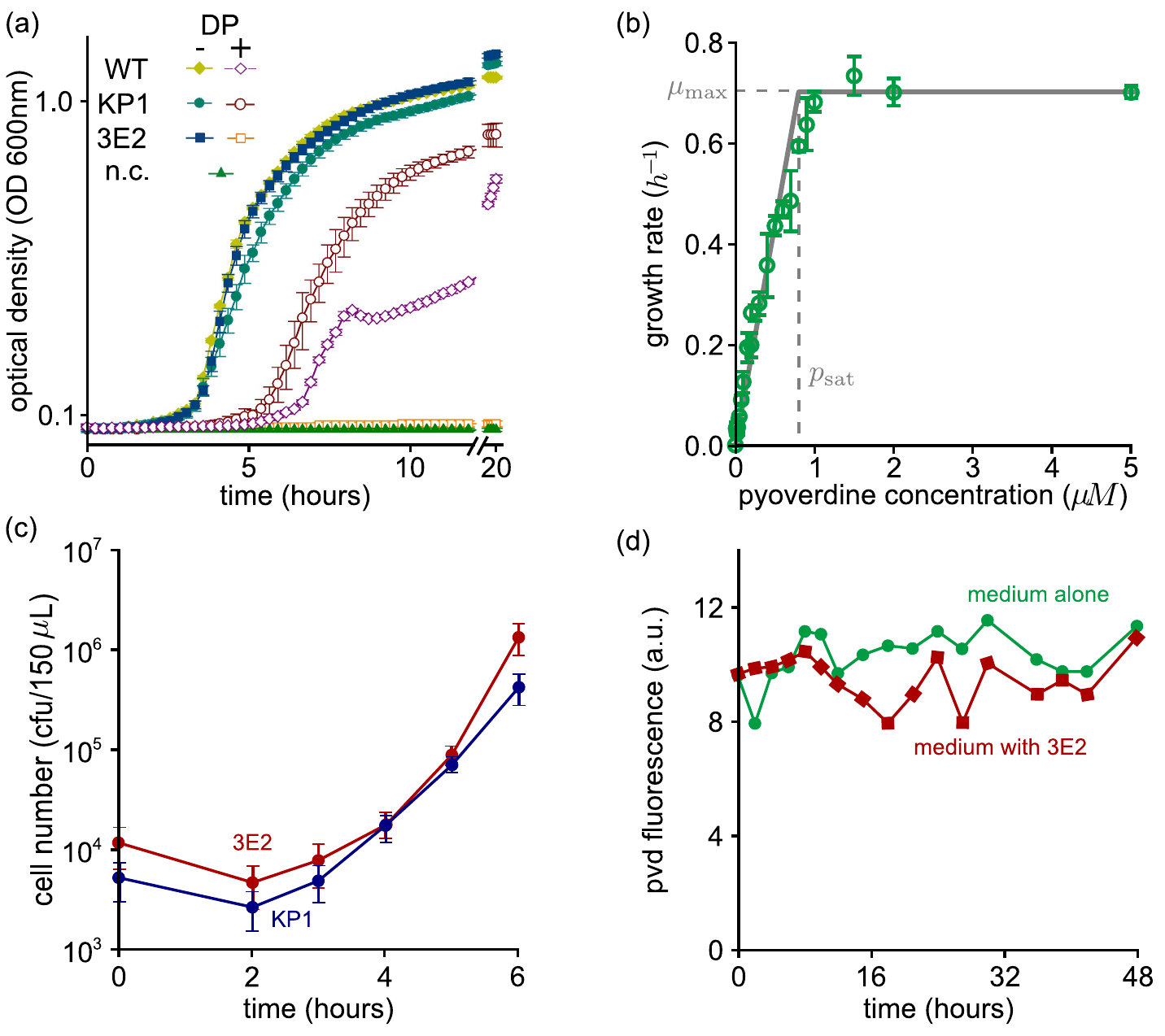}
\caption{
\textbf{Characterization of the fitness impact of pyoverdine}.
\textbf{(a)} In an environment with available iron (solid symbols), non-producer cells (strain 3E2, squares) grow as fast as the wild-type (WT, diamonds), and faster than constitutive producers (strain KP1, circles).
Under extreme iron limitation, pyoverdine is needed for growth: producers (and the wild-type) grow, whereas non-producers are unable to. negative control (n.c. , medium without \textit{P. putida} strains) is shown for comparison.
\textbf{(b)} Growth benefit from pyoverdine. Green dots represent the growth rate $\mu$ of non-producer cultures, with added pyoverdine.
The solid gray line represents the growth rate calculated using equation (\ref{eq:mu-p}) (with fitted maximal growth rate $\mu_\textnormal{max}= 0.878$ and saturation concentration $p_\textnormal{sat}=0.8$.
\textbf{(c)} Sharing and excludability.
In mixed populations, and under extreme iron limitation, producers (KP1, blue line) and non-producers (3E2, orange line) start growth together.
Since non-producers need pyoverdine to grow (see panel b), we conclude that pyoverdine is equally shared between the strains: it is non-excludable.
\textbf{(d)} Durability of pyoverdine.
Fluorescence of pyoverdine in growth medium, with and without non-producer bacteria.
Pyoverdine does not degrade spontaneously or through interaction with bacteria.
Reprinted from Ref. \cite{Becker_Wienand:2018}.
\label{fig:pvd-interaction-characterize}}
\end{figure}

To quantify the beneficial and cooperative role of pyoverdine, i.e. \/ its impact on population dynamics, one has to determine the metabolic load of pyoverdine production, its contribution to growth, its stability, and how evenly it is shared with other cells. 
As shown in Fig.~\ref{fig:pvd-interaction-characterize} (adapted from Ref.~\cite{Becker_Wienand:2018}) and discussed in the previous section, there is a cost associated with pyoverdine production. The costs were estimated by comparing growth of the constitutive producer with wild-type and non-producer under iron-replete conditions (Fig.~\ref{fig:pvd-interaction-characterize}a, solid symbols). Depending on growth conditions, the growth rate of the constitutive producer was 3-10\% lower than the growth rates of the two other strains.

In addition, the benefit of pyoverdine was quantified by growing non-producers alone in iron-depleted medium, in which cells need pyoverdine to grow.
As shown in Fig.~\ref{fig:pvd-interaction-characterize}b, the growth rate increases almost linearly with the concentration of added pyoverdine, and then sharply, at a threshold concentration $p_\textnormal{sat} \approx 1 \mu$M, levels off at a maximal growth rate $\mu_\textnormal{max}$, whose value depends on other provided culture conditions. 
To a very good degree of accuracy one has (see gray curve in Fig.~\ref{fig:pvd-interaction-characterize}b)
\begin{equation}
	\mu(p)
	=
	\begin{cases}
	\frac{\mu_\textnormal{max}}{p_\textnormal{sat}}, 
	& p<p_\textnormal{sat}
	\\
	\mu_\textnormal{max} 
	& p\geq p_\textnormal{sat}
	\end{cases}
	\; .
\label{eq:growth_putida}
\end{equation}

A good is dubbed \textit{excludable} if producers are able to prevent non-producers from accessing it~\cite{archetti_coexistence_2011}. This can be tested by cultivating producers and non-producers together in iron-depleted conditions and measure when each strain would start growing.
Figure \ref{fig:pvd-interaction-characterize}c shows that after an initial lag phase both strains initiate growth essentially at the same time. 
Since pyoverdine is absolutely necessary for growth in these conditions, this implies that both strains have equal access to it.
In other words, \textit{pyoverdine acts as a non-excludable good}.

Finally, Figure \ref{fig:pvd-interaction-characterize}d shows that pyoverdine is fluorescent for at least $\sim 48$ hours, and hence for all practical purposes it does not degrade spontaneously or through interactions with cells.
Therefore, while bacteria interact with pyoverdine, they do not appear to damage or degrade it.

Taken together, these observations characterise pyoverdine as a proper public good and show that producers incur a constant cost under given growth conditions.

Moreover, due to its very long lifetime, pyoverdine accumulates in the environment once produced. In Sections~\ref{sec:explicit_public_good} we consider the consequences of these properties of a public good on the emergence and stability of cooperative behavior. However, we first talk more about the general requirements for stable cooperation to emerge in structured microbial populations. 

\section{Selection and drift in structured (meta-)populations}
\label{sec:selection_drift}

\begin{figure}[tbh]
\centering
\includegraphics[width=0.9\textwidth]{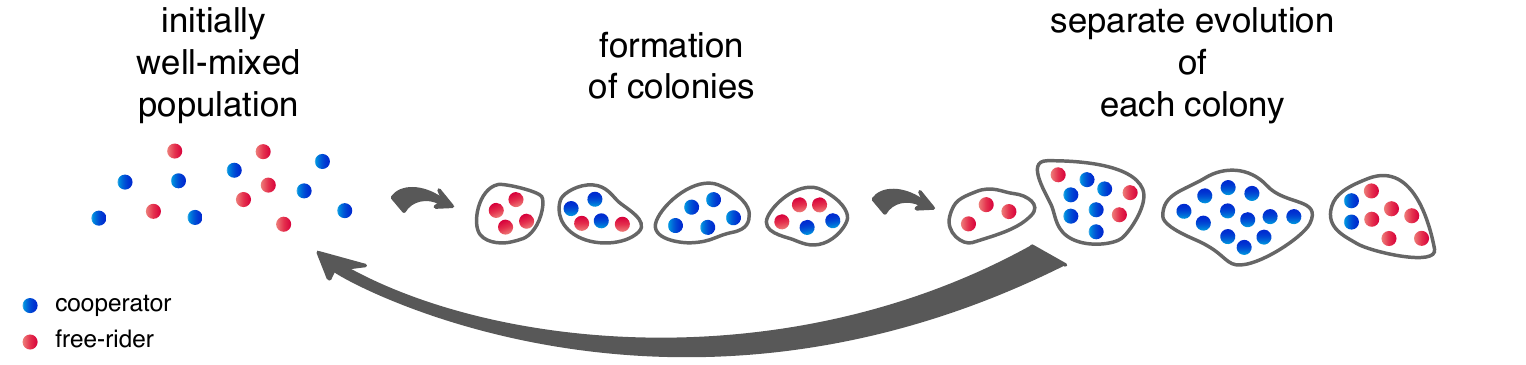}
\caption{{\bf A simplified life cycle of microbial populations.} This figure illustrates the simplified life cycle of a bacterial population consisting of cooperators (blue) and free-riders (red). There are three essential time periods: i) Initially, a well-mixed population is subdivided into groups of average size $n_0$ by some stochastic process. ii) Once these groups are formed, they independently and separately evolve according to a dynamics that has the following two key features. First, groups with a higher cooperator fraction grow faster as they contain more public good. Second, because free-riders do not have to provide the public good, they are growing faster than cooperators within each of these groups. iii) Finally, after some time $T$ the life cycle starts anew as all groups are merged into a single population. Adapted from Ref.~\cite{Melbinger:2015}.
}
\label{fig:cartoon}
\end{figure}

As we have outlined in Sections~\ref{sec:subseccooperationinthemicrobialworld}, microbes live in structured populations where they exhibit periods of strong growth and periods of dynamic restructuring. 
Theoretical studies suggest that such a population structure might promote the stability of cooperation (Section~\ref{sec:factors_cooperationinmicrobialcooperation}).
Indeed, experimental studies under well controlled experimental conditions have confirmed this by investigating the consequences of different population structures for a range of bacterial sytems (Section~\ref{sec:experimental_model_systems}). To conceptually analyse microbial population structures consider a life cycle model consisting of three steps~\cite{Cremer_Melbinger:2011} as illustrated in  Fig.~\ref{fig:cartoon}.

\begin{enumerate}
\item[(i)] \emph{Group-formation step:}  initially, a well-mixed population consisting of cooperators and non-cooperating \textit{defectors} (\emph{free-riders}) is randomly assorted into different groups (colonies). 
\item[(ii)] \emph{Group-evolution step}: next, each of the so formed groups evolves independently and separately following a growth law that is specific to the  microbial system under consideration. 
\item[(iii)] \emph{Group-merging step}: the life cycle starts anew after some time $T$ when the colonies are merged together into one population. 
\end{enumerate}

This scheme closely follows controlled laboratory experiments~\cite{Chuang:2009,Chuang:2010}, see the discussion in Section~\ref{sec:experimental_model_systems}. The idealized scheme as well as these experiments are obviously simplifications of the complex arrangement, growth, and restructuring patterns which are happening in the wild. However, they both capture major aspects of microbial growth. In particular, they capture the key feature of regularly occurring population bottlenecks, namely rearranging local sub-populations (colonies) whose initial population size consists of only a few individuals; see also the discussion in Section~\ref{sec:cooperationinmicrobialworld}. Furthermore, as the random assortment of individuals constitutes a worst-case scenario for cooperation (no preferential assortment of cooperators with other cooperators), it allows to study the minimal conditions for the possible onset of cooperation in microbial populations starting from a single cooperative mutant~\cite{Melbinger:2015}. 


In the following subsections, we will first discuss the \emph{group formation} and \emph{group evolution steps} in systems where random drift dominates and selection effects can be neglected. 
Next, we will review recent theoretical and experimental results on the combined effect of assortment noise, demographic noise, and selection pressure by public good production.
Finally, in the last subsection, we give an overview on theoretical results showing that cooperation can be maintained in systems running repeatedly through a life cycles where all of the above three steps are repeated cyclically.

\subsection{Random drift in growing bacterial populations}
\label{sec:random_drift_bact_pop}

In Section \ref{sec:fluctuations} we discussed the role of (demographic) noise in populations of constant size. 
The main point was that the strength of demographic noise scales inversely with the population size $N$. As a consequence, when the population size is changing, non-selective effects are dominant over selective effects when the population size is small. However, these non-selective effects become more and more sub-dominant with increasing population size. This raises the important question of how the interplay between demographic noise and population growth affects the composition of initially small populations and how these results differ from systems with large populations of fixed size. In this section, we review recent results for systems which are neutral, i.e. where selection is completely absent (no public good production and strains show similar growth behavior). This analysis will also be important when we consider the case with selection.

\begin{figure}[tb!]
\centering
\includegraphics[width=0.5\textwidth]{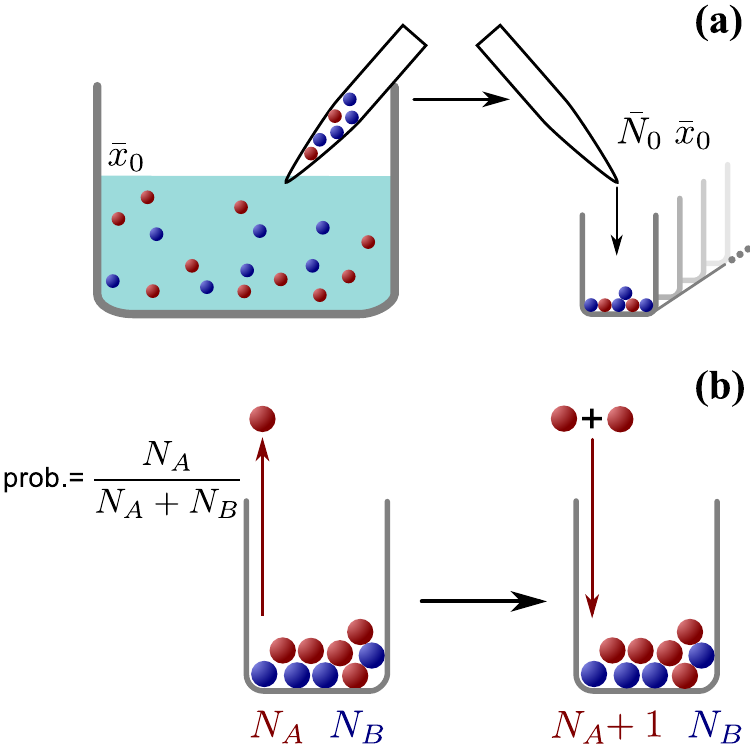}
\caption{
\textbf{Urn sampling and growth}.
\textbf{(a)} \textit{Schematic illustration of the random initial conditions}. Sub-populations are initiated by repeatedly taking small volumes, containing a small number of cells, from a large reservoir containing a dilute mixture of two strains of bacteria, $A$ and $B$, indicated by marbles of different color. The fraction of $A$ strains is given by $\bar x_0$. This sampling process implies that the initial number of individuals in the sub-populations follows a Poisson distribution, as specified in the main text. 
\textbf{(b)} \textit{Illustration of the P\'olya urn model}. The urn model is a stochastic process where marbles of two different colors are repeatedly drawn at random with a probability given by their relative abundances. After drawing they are returned back together with a second marble of the same color. In order to create a process exhibiting exponential growth one has to choose the waiting time between successive iterations to be exponentially distributed (\textit{Poissonization}). Adapted from Ref.~\protect\cite{wienand_non-selective_2015}.
}
\label{fig:setup}
\end{figure}

Pyoverdine production by \textit{P. putida} KT2440 was recently used as a model to study such neutral dynamics~\cite{wienand_non-selective_2015}. Co-cultures of \textit{P. putida} KT2440 (wild type strain) and an isogenic mutant (non-producer, 3E2) were mixed, diluted to yield different sub-populations with Poisson distributed initial composition, and grown under high iron conditions. Under these conditions, KT2440 is not producing pyoverdine and both strains show similar growth behavior (see Section \ref{sec:pyoverdine}). With this setup, the dynamics of different sub-populations with a random distribution of initial cell numbers $N_0$ and producer fraction $x_0$ can be studied, as illustrated in Fig.~\ref{fig:setup}. 
The experimental setting just described is equivalent to the classic \textit{P\'olya urn model}~ \cite{eggenberger_uber_1923} if one identifies marbles in the urn with bacteria in a well-mixed well.
In its most basic formulation~\cite{eggenberger_uber_1923}, sketched in Fig.~\ref{fig:setup}b, this model describes an urn containing $N$ marbles of two different colours.
At each step, a marble is drawn at random, and then placed back, alongside another one of the same color.
This way the urn grows in size at each step.
Each marble is equally likely to be drawn, therefore the probability of extracting a marble of a certain color is equal to the relative abundance of marbles of that color in the urn, for example, $\text{Prob}
	\{ \text{``extract red''} \} 
	= 
	N_A/(N_A+N_B)$. 
Every extraction adds a marble to the same color to the urn, thus increasing the chance of drawing further marbles of that same color in the future.
Therefore, the P\'olya urn is a \textit{self-reinforcing} (or \textit{auto-catalytic}) \textit{process}~\cite{brian_arthur_path-dependent_1987,pemantle_survey_2006}.
One could intuitively expect this positive feedback to amplify each small initial fluctuation.
As more and more marbles of the same color are drawn, it becomes more and more likely to extract that color, and eventually the urn would come arbitrarily close to being homogeneous.
In biological terms, this corresponds to an almost fixated population, analogous to the result of genetic drift.
At the same time, however, the increasing number of marbles decreases the impact of individual fluctuations.
Each added marble changes the proportions in the urn less than the previous one.

\begin{figure}[tbh]
\centering
\includegraphics[width=0.6\linewidth]{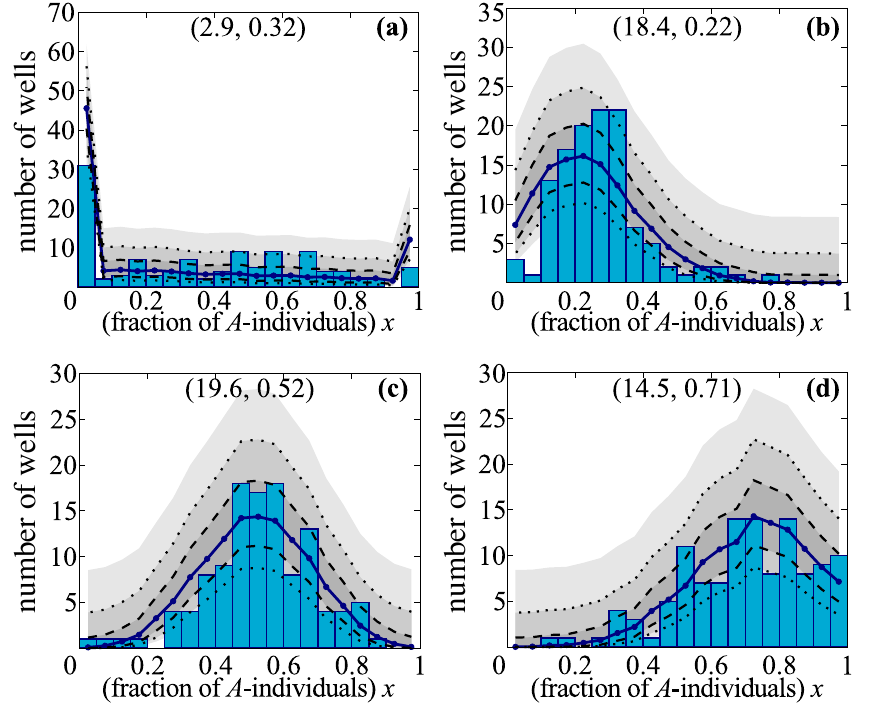}
\caption{
\textbf{Steady-state distributions of population composition} are shown for both experiments (bars) and theoretical model for a series of different initial conditions. The averages $x_0$ and $N_0$ determined from the experiment (using 120 independent wells) were used to initialize the stochastic simulations for a set of $120$-well ensembles. The blue solid line indicates that average theoretical distribution obtained from averaging the histograms obtained from these set of ensembles after growth. The shaded areas show confidence intervals: 68\% (between dashed lines), 95\% (between dotted lines) and 99.73\%. The parameter values, also indicated in the graph are: \textbf{(a)} $\bar N_0 = 2.9$, $\bar x_0 = 0.32$; \textbf{(b)} $\bar N_0 = 18.4$, $\bar x_0 = 0.22$; \textbf{(c)} $\bar N_0 = 19.6$, $\bar x_0 = 0.52$;
\textbf{(d)} $\bar N_0 = 14.5$, $\bar x_0 = 0.71$.
Adapted from Ref.~\protect\cite{wienand_non-selective_2015}.
\label{fig:polya-experiments}}
\end{figure}

Mathematical and experimental studies \cite{wienand_non-selective_2015} using different initial conditions (Fig.~\ref{fig:polya-experiments}), i.e.\/ different combinations of the initial average population size $\bar N_0$ and composition $\bar x_0$,
have lead to the following insights: 
(i) Fixation of a population is not due to fixation during population growth but simply a consequence of the initial sampling process. For small average initial population size or compositions close to $x=0$ or $x=1$, this effect is strongest. Here a large fraction of the wells initially contain cells of strain $A$ or $B$ only; see Fig.~\ref{fig:polya-experiments}\textbf{(a)} and Fig.~\ref{fig:polya-experiments}\textbf{(d)}.
(ii) The composition of the population approaches a random stationary limit, i.e. the probability distribution for the population composition converges to limit distributions (Fig.~\ref{fig:polya-experiments}), which are distinct from Kimura's result for  populations with a constant population size~\cite{KimuraFixation}. 


\subsection{Evolutionary game theory in growing populations}
\label{sec:egt_growing_populations}

What happens if the dynamics is not neutral but metabolic costs for the production of the public good are significant?
In this section, we will discuss a recently introduced generalized game theory framework which enables one to consider the combined effect of internal population dynamics and changes in population size~\cite{Melbinger:2010, Cremer_Melbinger:2011}.
In this theory one starts with a stochastic model where individual agents reproduce or die with rates that depend on the state of the population, instead of by winning a `tooth-and-claw' battle with an opponent where one's survival directly results in the death of the opponent. 
This has conceptual advantages as it allows a more biological interpretation of evolutionary dynamics than common formulations using the Fisher-Wright or Moran process~\cite{Ewens, Blythe:2007, Moran}. 

In the simplest setting, one considers an ensemble of subpopulations with only two different traits: 
each individual may either be a  cooperator and defector (free-rider)~\cite{Melbinger:2010,Cremer_Melbinger:2011}. The state of each sub-population $i$ is defined by the number of cooperators, $N_C^{(i)}$, and defectors, $N_D^{(i)}$, or equivalently, the total number of individuals, $N^{(i)} = N_C^{(i)} + N_D^{(i)}$, and the fraction of cooperators, $x_i= N_C^{(i)} / N^{(i)}$. 
The stochastic population dynamics is defined by birth ($A \to 2A$) and death ($A\to\emptyset$) rates for each trait $A \in \{ C, D \}$ 
\begin{subequations}
\begin{align}
	\Gamma_{A \to 2A} 
	&= 
	\mu (x) \, f_A^{} (x) \, N_A^{} 
	\; , \\
	\Gamma_{A\to\emptyset}
	&= d(N) \, N_A^{}
	\; .
\label{eq:rates}
\end{align}
\end{subequations}
The quantity $f_A^{} (x)$ denotes the fitness (growth advantage or disadvantage) for each of the strains, and in the simplest case is given by $f_C = 1-s$ and $f_D = 1$, where $s$ denotes the cost for cooperation, e.g.\/ by production of a common good like siderophores.
Importantly, cooperation positively affects the whole population by increasing its global fitness. 
Hence, one assumes that the population growth rate $\mu (x)$ increases with the cooperator fraction $x$.
In its simplest form it is often taken as linear $\mu (x) = 1 + p x$ where $p$ is an effective parameter characterising the growth advantage of populations containing a larger amount of cooperators. 
 For instance, for \textit{P. aeruginosa}, as discussed in Section \ref{sec:pyoverdine}, the iron uptake, and hence the birth rates, increase with a higher siderophore density and therefore with a higher fraction of cooperators, see Fig.~\ref{fig:pvd-interaction-characterize}b. In general, $\mu (x)$ may also be a nonlinear function of $x$, as e.g. found in Ref.~\cite{Gore:2009}. The specific choice depends on the particular biological system under consideration. The per capita death rate $d(N) = N/K$ models logistic growth with carrying capacity $K$ accounting for limited resources. 
 
\begin{figure}[t]
\begin{center}
\includegraphics[width=\textwidth]{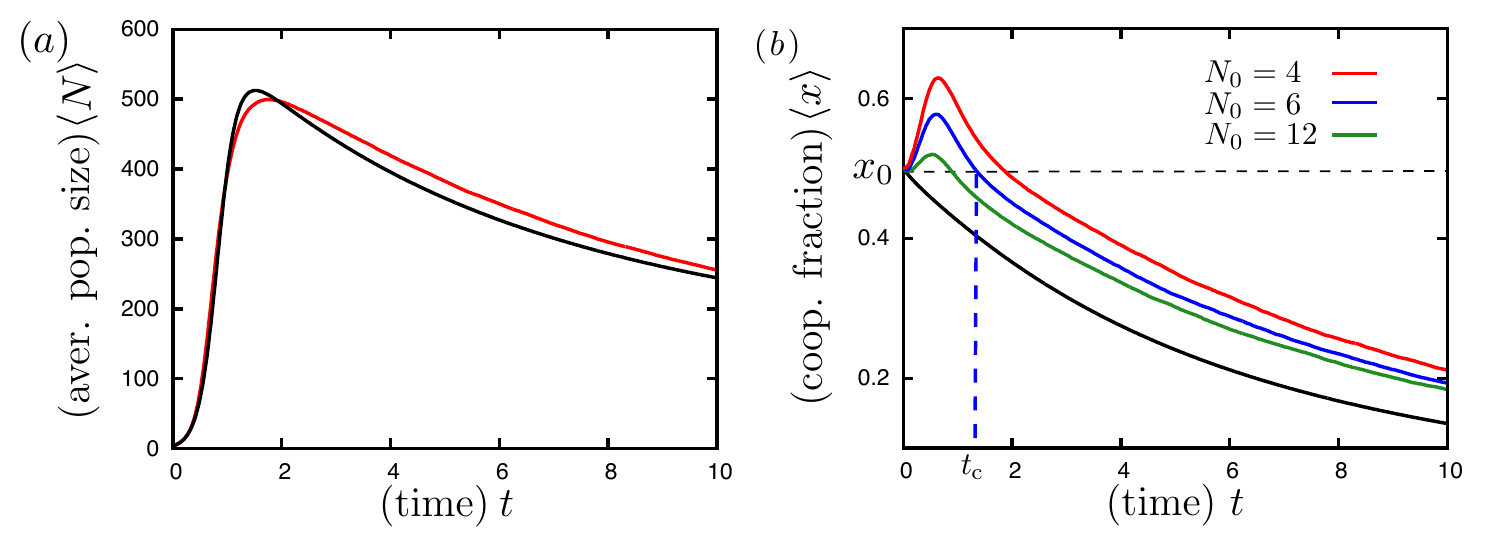}
\end{center}
\caption{
\textbf{Transient increase in cooperation.} The figures show the temporal evolution of the average population size $\langle N\rangle$ and the cooperator fraction $\langle x \rangle=\sum_iN^{(i)}_{C}(t)/\sum_i N^{(i)}(t)$ 
averaged over the ensemble of all subpopulations.
{\bf (a)} \textit{Average population size.} The results obtained from the stochastic simulations for $N_0 = 4$ (red line) agree well with the approximation obtained from the deterministic equations (black line). The initial rise and subsequent decline of the population size is due to logistic growth and a decrease in carrying capacity due to selection, respectively.
{\bf (b)} \textit{The fraction of cooperators.} As described in the main text, the initial increase results from an asymmetric amplification of fluctuations, and the subsequent decline is due to selection. The cooperation time $t_C$ is defined as the initial time period where the cooperator fraction $x$ lies above its initial value $x_0$. This effect is stochastic and not accounted for by a deterministic theory (black line). Initial population sizes are indicated in the graph. Other parameters used in the simulations are $s=0.1$ and $p=10$. Adapted from Refs.~\cite{Melbinger:2010, Cremer_Melbinger:2011}. 
}
\label{fig:unfrozen}
\end{figure}

\textit{Deterministic population dynamics.} Neglecting any kind of demographic noise, the time evolution of the population would be given by a set of deterministic equations -- also called mean-field equations -- for the average population size, $N$, and the average cooperator fraction, $x$, respectively:
\begin{subequations}
\begin{align}
	\partial_t N
	&=
	\left(\mu (x)-\frac{N}{K}\right)N
	\label{eq:pdN} 
	\, , \\
	\partial_t x
	&=-s \, \mu (x) \, x(1-x) 
	\, .
\end{align}	
\label{eq:deterministic_transient}
\end{subequations}
Here, the dynamics for the internal composition of the population reduces to the common game theory scenario~\cite{Moran, Nowak:2004, Traulsen:2005}, where a changing population size is immaterial to the evolutionary outcome of the dynamics~\cite{Blythe:2007}. 
Therefore, the average fraction of cooperators $x$ would decrease monotonously in time as shown by the black curve in Fig.~\ref{fig:unfrozen}b.
The coupling between population dynamics and internal dynamics merely leads to an overshoot in the population size $N$, as shown in Fig.~\ref{fig:unfrozen}a.

\textit{The role of demographic noise.} 
The internal dynamics becomes qualitatively different if demographic noise is accounted for~\cite{Melbinger:2010, Cremer_Melbinger:2011}. 
In principle, there are two possible sources of noise: extrinsic noise due to random assortment of individuals into subpopulations as discussed in the previous section and intrinsic noise due to demographic noise. 
Here, we focus on the latter. 
However, similar results are obtained when considering extrinsic noise or both types of randomness~\cite{Cremer_Melbinger:2011, Melbinger:2015}.
We will come back to this later, and for now consider a (large) set of subpopulations, all of the same initial size, $N_0$, and the same initial internal composition (initial cooperator fraction), $x_0$. 
As shown in Fig.~\ref{fig:unfrozen}b, one finds a transient increase in the average fraction of cooperators.
This shows that demographic noise during population bottlenecks can lead to a finite time period $t_C$ where the cooperator fraction rises above its initial value $x_0$ such that during this time period the selection disadvantage of cooperators is overcome.
Since the strength of demographic noise scales as $\sqrt{1/N}$ ~\cite{Blythe:2007, Kimura, Cremer:2009}, the duration of the corresponding cooperation time $t_C$ strongly depends on the initial population size $N_0$~\cite{Melbinger:2010, Cremer_Melbinger:2011}. 
Importantly, the transient increase in cooperator fraction is a genuine stochastic effect which is \emph{not} captured by the deterministic equations, Eq.~\eqref{eq:deterministic_transient}.
It is caused by the amplification of stochastic fluctuations during the initial phase of the population dynamics where the population size is still small. 
During this phase, demographic noise is strong and cooperator fractions will be widely different in the various subpopulations. 
There is an asymmetry in the amplification of stochastic fluctuations because of the coupling between population growth and its internal composition. While an additional cooperator amplifies the growth of a population, an additional defector has just the opposite effect.
This implies an asymmetric amplification of demographic favouring those subpopulation in an ensemble that contain a larger cooperator fraction~\cite{Melbinger:2010}. 
As a consequence, the ensemble average 
\begin{equation}
	x(t)
	= 
	\frac{\sum_{i}^{}N_{C}^{(i)}(t)}
	     {\sum_i^{} N_{}^{(i)}(t)}
	\, ,
\end{equation} 
i.e. the mean fraction of cooperators obtained when averaging over different realisations $i$, may show an increases with time. 
This effect is only transient as with growing population size the selection pressure towards more defectors becomes dominant~\cite{Melbinger:2010, Cremer_Melbinger:2011}. 

Taken together, the main insight gained from these studies of the prisoner's dilemma is that demographic noise can result into a fluctuation-induced transient increase of cooperation. 
An essential prerequisite for this stochastic effect is the presence of a positive correlation between global population fitness and the level of cooperation. 

\subsection{The role of explicit public goods}
\label{sec:explicit_public_good}
 
Theoretical models of public good exchange, like the one we discussed in the previous section, typically leave social interactions implicit and formulated in terms of game-theory models~\cite{Frey:2010, Gore:2009, Hofbauer, Reichenbach:2007, Reichenbach_Mobilia:2007, Reichenbach:2008} or inclusive fitness models~\cite{Griffin_West:2004, Kummerli_Gardner:2009, Nowak:2006, Julou:2013}. 
As discussed in Section \ref{sec:evolutionary_game_theory}, a paradigmatic example is the prisoner's dilemma game~\cite{axelrod-1981-211, Damore_Gore:2012, frank_foundations_1998, Hamilton:1964}.
While these approaches provide important conceptual insights, it is now becoming increasingly clear that a more detailed molecular view on public goods is necessary to quantitatively understand the population dynamics of bacterial systems.  
Possible effects include the diffusion of public goods~ \cite{Julou:2013, Kummerli:2014, Dobay:2014}, a regulatory role of public good production~\cite{MacLean:2010}, the 
interference of different public goods with each other \cite{inglis_presence_2016}, and the function of public goods in inter-species competition~\cite{Niehus:2017}.

\begin{figure}[!t]
\vspace{0.5truecm}
\center
\includegraphics[width=1\linewidth]{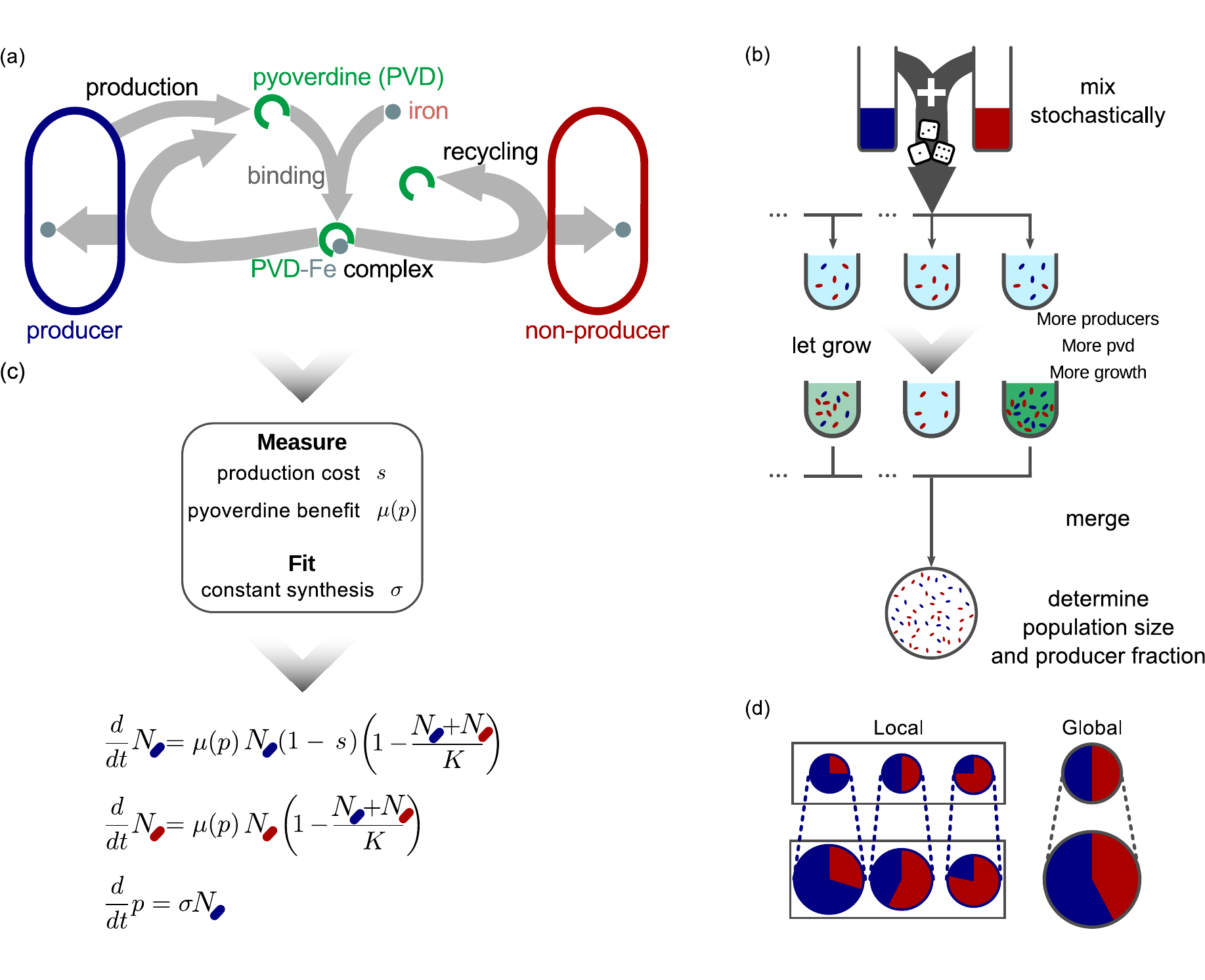}
\caption{
\textbf{Quantitative analysis of an experimental model population.}
\textbf{(a)} \textit{Sketch of the interactions in the bacterial model system}: Producers secrete pyoverdine, which binds to iron in the medium.
All cells (regardless of the strain) absorb the complex into their periplasm, where it is separated: iron is transported inside the cell, whereas pyoverdine is secreted back in the environment.
\textbf{(b)} \textit{Sketch of the metapopulation experiment}: Pure cultures are mixed in stochastic proportions to inoculate the wells of a well-plate.
This ensures that each well contains a stochastic fraction of producers.
Populations containing more producers (blue) benefit from their more abundant pyoverdine and grow faster.
\textbf{(c)} \textit{Road map to establish a mathematical model}: The main parameters of the interaction, such as costs and benefits of pyoverdine, are quantified experimentally as discussed in Section~\ref{sec:pyoverdine}.
A single parameter (the synthesis rate of pyoverdine) is left as a fit parameter.
At given intervals, samples were taken from each well and then merged to measure both the average population size and the global producer fraction.
\textbf{(d)} Sketch of Simpson's paradox (adapted from Ref. \cite{Chuang:2009}): Producer fractions within each population (blue portions of the left pie charts) always decrease; however, as long as more producing populations grow larger (larger pies), the global producer fraction across the ensemble (right pie chart) may increase.
\label{fig:pyov-gfx-abs}}
\end{figure}

Experimental model systems like \textit{Pseudomonas} populations allow one to study interactions between bacteria that involve the exchange of a public good, specifically the iron-scavenging compound pyoverdine. Based on the experimental quantification  of costs and benefits of pyoverdine production (see Section \ref{sec:pyoverdine}) it becomes possible to build \textit{quantitative} theoretical models of population dynamics that explicitly account for the changing significance of accumulating pyoverdine as a chemical mediator of social interactions. Here, we review recent progress made in understanding the role of public goods, particularly by studying pyoverdine production in \textit{Pseudomonas putida} as an experimental model system~\cite{Becker_Wienand:2018}.
As discussed in detail in Section \ref{sec:pyoverdine} and as sketched in Fig.~\ref{fig:pyov-gfx-abs}(a), a single public good (pyoverdine) mediates cell-cell interactions. 
To minimize complexity, the model system contains two engineered strains, a strain (KP1) which constitutively produces pyoverdine, and a non-producer strain (3E2). The experimental setup, illustrated in Fig.~\ref{fig:pyov-gfx-abs}(b), considers a metapopulation where individual subpopulations were initialized as stochastic mixtures of the two strains. The metapopulation consisted of a 96-well plate with each well representing a subpopulation (inoculated with about $10^4$ cells).
The stochastic sampling of the strain composition in the subpopulations (wells) was chosen such that it mimics the characteristic variability of small populations.

Figure~\ref{fig:pyov-gfx-abs}b illustrates that samples were taken from each well at given set of time points $t$, and then merged to determine the average cell number 
\begin{equation}
	\langle N \rangle (t)
	=
	\frac{1}{M}
	\sum_{i=1}^M 
	\left(
	N_C^{(i)}(t) + N_D^{(i)}(t)
	\right)
\end{equation}
and the mean producer fraction across the metapopulation
\begin{equation}
	\langle x \rangle (t)
	=
	\frac{\sum_{i=1}^M N_C^{(i)}(t)}
	     {\sum_{i=1}^M  \left(
	                    N_C^{(i)}(t) + N_D^{(i)}(t)
	                    \right)}
	\, .
\end{equation}

\textit{Mathematical model.} Building a 
mathematical model of this experimental model system has to be based on the known characteristics of the experimental model system; see Fig.~\ref{fig:pyov-gfx-abs}c.
In particular, one must take into account the availability of pyoverdine and its effect on growth:
(i) pyoverdine molecules find and bind an iron atom as soon as they are released; 
(ii) cells absorb the bound pyoverdine-iron complex at a constant rate; 
(iii) cells try as much as possible to maintain their internal iron concentration constant (iron homeostasis); 
(iv) up to some saturation concentration, iron is the only factor limiting growth.
Mathematically, this can can be integrated into a growth rate of the form (see Eq.~\ref{eq:growth_putida})
\begin{equation}
	\mu(p)
	=
	\mu_\textnormal{max} \,
	\min
	\left(
	\frac{p}{p_\textnormal{sat}},1
	\right)
	\, ,
\label{eq:mu-p}
\end{equation}
where $p$ denotes the concentration of pyoverdine, and $p_\textnormal{sat}$ represents the concentration at which iron availability stops being growth-limiting.
Equation (\ref{eq:mu-p}) reflects the linear dependence of $\mu(p)$ on the pyoverdine concentration $p$ as observed in experiments (see Fig.~\ref{fig:pvd-interaction-characterize}b in Section \ref{sec:pyoverdine}).
At the deterministic level\footnote{Experimental population started with ${\sim} 10^3$ individuals and grew to a final size of ${\sim} 10^6$, justifying such deterministic description neglecting demographic noise.} of description this implies the following growth relations for the amount of producer cells $N_C$ and non-producers $N_D$ in a population \cite{Becker_Wienand:2018}.
\begin{subequations}
\begin{align}
	\partial_t N_C 
	&= 
	\mu(p) \, N_C \,
	(1-s) \, 
	\left(
	1-\frac{N_C+N_D}{K}
	\right)
	\, ,
	\label{eq:pvd-pop-unscaled}
	\\
	\partial_t N_D 
	&= 
	\mu(p) \, N_D \,
	\left(
	1-\frac{N_C+N_D}{K}
	\right)
	\, , 
\end{align}	
\end{subequations}
where producers are assumed to grow slower by a factor $1-s$ because of the cost of production.
These growth relations account for finite nutrient supply by the carrying capacity $K$ (logistic growth). 
When the population exhausts the resources in the environment (determined by a carrying capacity $K$), cells end growth by entering a dormant state (see Section~\ref{sec:popgrowth}).
Finally, the amount of pyoverdine in the bacterial system is not constant, as producers synthesize pyoverdine at a constant per-capita rate $\sigma$: $\partial_t p = \sigma N_C$.

It is instructive to analyse the time evolution of the population size and composition of the population separately. 
To this end, it is convenient to consider the rescaled variables $n:=(N_C+N_D)/K$ (how big the population is, compared to the maximal size the environment allows), $x:=N_C/(N_C+N_D)$ (the fraction of producers in the population), and $v:=p/p_\textnormal{sat}$ (how far the pyoverdine concentration is from saturation concentration).
Moreover, it is convenient to rescale time in terms of the maximal growth rate $\mu_{max} = \mu(p_{sat})$ and redefine $\mu(v) = \min(v,1)$.
Altogether, the population dynamics if described by the following rescaled equations:
\begin{subequations}
\begin{align}
	\partial_t n 
	&= 
	n \, \mu(v) \, 
	(1-sx) \, (1-n)
	\, ,
	\\
	\partial_t x 
	&= 
	-s \, \mu(v) \, 
	x \, (1-x) \, (1-n)
	\, ,
	\label{eq:population-equations}
	\\
	\partial_t v 
	&= 
	\alpha \, n \, x
	\, .
\end{align}
\end{subequations}
Here, $\alpha$ defines a (dimensionless) \textit{accumulation parameter}
\begin{equation}
	\alpha
	:=
	\frac{\sigma \, K}
	     {p_\textnormal{sat} \, \mu_\textnormal{max}}
	\, ,
\end{equation}
which measures how fast pyoverdine accumulates compared to population growth.
High $\alpha$ corresponds to fast public good production, rapidly leading to high pyoverdine concentration; low $\alpha$ means that cells reproduce much faster than they can synthesise pyoverdine, which limits growth for long times.

\textit{Transient increase in producer fraction.} 
In the experiments~\cite{Becker_Wienand:2018} the strains were randomly assorted to the subpopulations (wells of a 96-well plate) as previously introduced. Starting with a well-mixed population containing a cooperator fraction $x_0$ one forms  a set of $M$ groups (wells), where each of these wells contains a randomly chosen cooperator fraction $x_0$. In these experiments the population sizes were relatively large, starting already with around $10^3-10^4$ individuals and then growing to sizes with up to $10^7$ cells. Hence, stochasticity in the initial size is low such that one can assume that all populations have approximately the same size. The statistical distribution of the cooperator fractions was determined from the experimental data~\cite{Becker_Wienand:2018}.

\begin{figure}[!t]
\centering
\includegraphics[width=\linewidth]{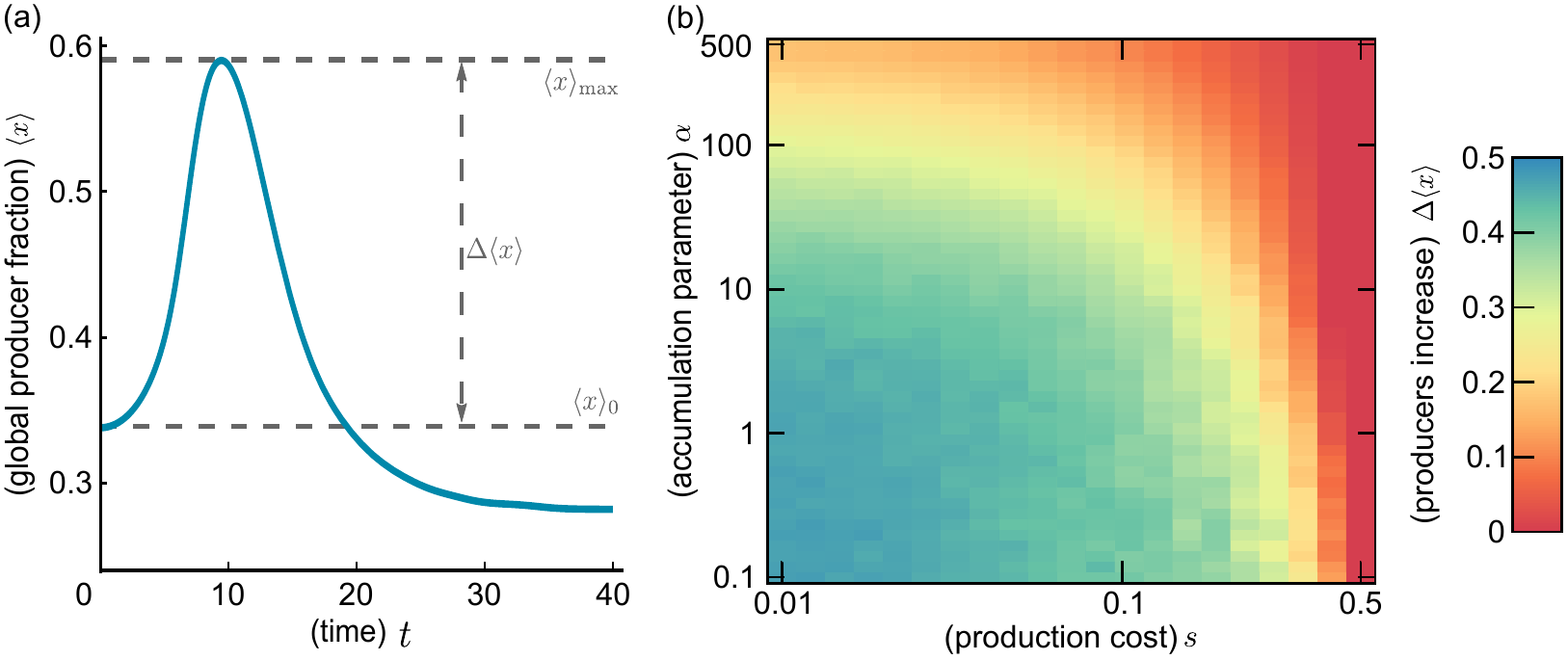}
\caption{
\textbf{Transient increase in global producer fraction.}
\textbf{(a)} Global producer fraction $\langle x\rangle$ as a function of time obtained in simulations using parameter values $\alpha=200$, $s=0.05$, $n_0=10^{-3}$.
Solving the deterministic equations for the population dynamics and averaging over the ensemble of initial conditions one observes first an increase in the global producer fraction, which peaks at a value $\langle x \rangle_\textnormal{max}$, then decreases (typically below its initial value) and eventually saturates to a stationary long-term value.
\textbf{(b)} Magnitude of the maximum producer increase $\Delta \langle x \rangle$ (colour code) as function of the pyoverdine accumulation (parameter $\alpha$) and the production cost $s$.
Increasing the production cost (horizontal axis, logarithmic scale) burdens producers, thus curtailing the increase. The main advantage of producer populations is rooted in their faster growth.
Higher accumulation of pyoverdine (parameter $\alpha$, vertical axis), however, leads to a faster saturation of public good levels, reducing the marginal benefit of pyoverdine. It also leads to low-producer sub-populations growing sooner. Both effects contribute to reduce the amplitude of the increase.
\label{fig:pyov-phasediag}}
\end{figure}

A typical trajectory resulting from the solution of the mathematical model, Eq.~\eqref{eq:population-equations}, using the experimentally determined initial conditions, is shown in Fig.~\ref{fig:pyov-phasediag}.
One finds that over a broad parameter range, the population average in the cooperator fraction, $\langle x \rangle$, shows a transient increase, i.e.\/ it initially increases, peaks at a value $\langle x \rangle_\textnormal{max}$ before decreasing, and finally saturating to some stationary value.
This characteristic profile can be rationalised as follows.  The initial growth rate of a subpopulation strongly correlates with the fraction of producers it contains, thereby driving the increase in global producer fraction, in accordance with the Price equation \cite{Melbinger:2010, Price:1970, price_extension_1972}.
Over time, however, pyoverdine accumulates in each subpopulation.
As a consequence the benefit of pyoverdine to individual cells will saturate, which in turn leads to a reduction in the advantage producer-rich populations which they had during initial stages.
At the same time, subpopulations with few producers also accumulate enough of the public good to start growing.
Eventually, producer-rich populations enter the dormant phase, and producer-poor ones catch up with their size.
This implies that the increase in the global producer fraction is only a transient phenomenon.

\begin{figure}[!t]
\center\includegraphics[width=\linewidth]{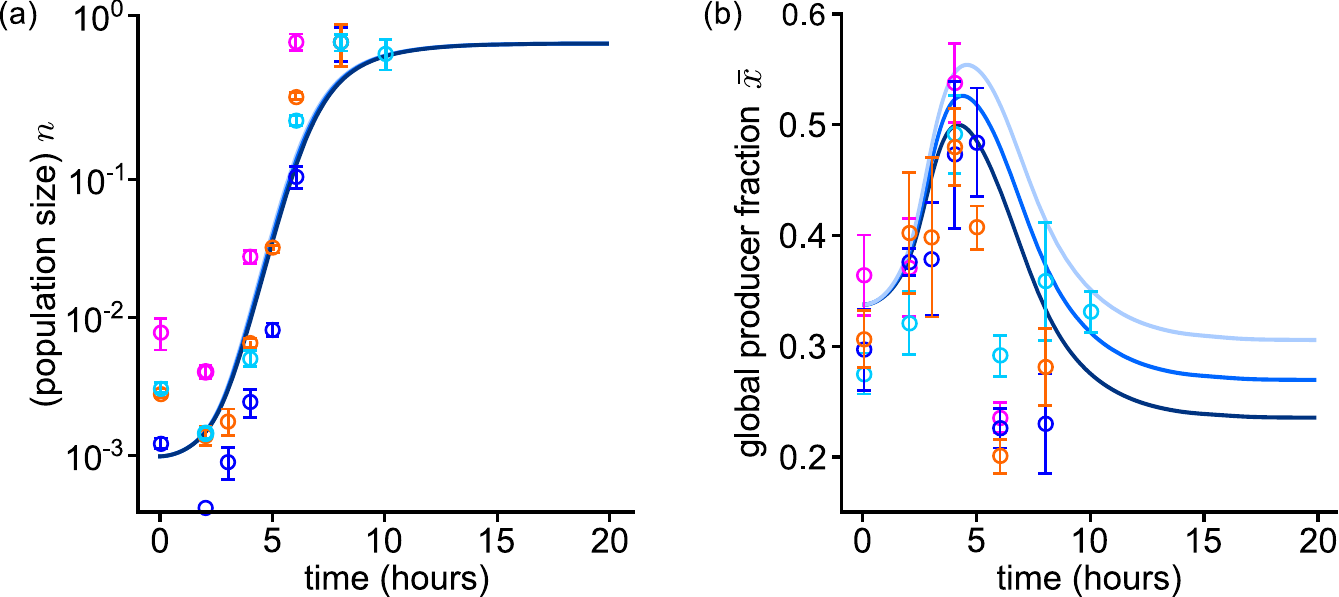}
\caption{
\textbf{Comparison of theory and experiments}. 
The average population size $n$, rescaled to the final yield (or carrying capacity, (\textbf{a}), and the fraction of cooperators (\textbf{b}) observed experimentally~\cite{Becker_Wienand:2018} is compared with the numerical solution of Eq.~\ref{eq:population-equations} (solid lines) for a set of values for the production cost $s\in \{0.03,0.05,0.07\}$; the darker shades of blue indicate higher values of $s$. The accumulation parameter $\alpha=200$ was obtained by fitting the numerical results to the experimental data. Experimental error bars are the standard deviations of three to five replicates of the experiment.
Reprinted from Ref. \cite{Becker_Wienand:2018}.
\label{fig:pvd-comparison}}
\end{figure}

The amplitude of the transient increase $\Delta \langle x\rangle = \langle x\rangle_\textnormal{max}-\langle x\rangle_0$ depends on both the production cost $s$ and the accumulation parameter $\alpha$, as shown in Fig.~\ref{fig:pyov-phasediag}b.
Clearly, a higher cost $s$, i.e.\/ a heavier burden on producers, leads to less of an increase in global producer fraction.
The accumulation parameter $\alpha$ also reduces $\Delta \langle x\rangle$.
For $\alpha \rightarrow0$, in fact, pyoverdine is produced much slower than the population growth.
Scarce pyoverdine strictly limits growth for several generations, during which only producer-rich populations can grow appreciably.
Moreover, saturation of the public good only happens rather later, allowing producer-rich populations to maintain their growth advantage for long times, which leads to a high increase in producer fraction.
In contrast, for $\alpha \gg 1$, production and accumulation of pyoverdine occur faster than cell replication.
A few producers thus suffice to rapidly accumulate enough public good to reach saturation.
Hence, producer-rich populations only briefly have an advantage, and the resulting increase is lower.

Comparing the simulated transient increase with experiments confirms the approach, see Fig.~\ref{fig:pvd-comparison}. The only fit parameter in the comparison between experiments and simulations was the accumulation parameter $\alpha$; all other model parameters were inferred from the experimental data~\cite{Becker_Wienand:2018}.

\subsection{Microbial life cycles: maintenance and evolution of cooperation}
\label{sec:life_cycles}

\begin{figure}[!t]
\centering
\includegraphics[width=\textwidth]{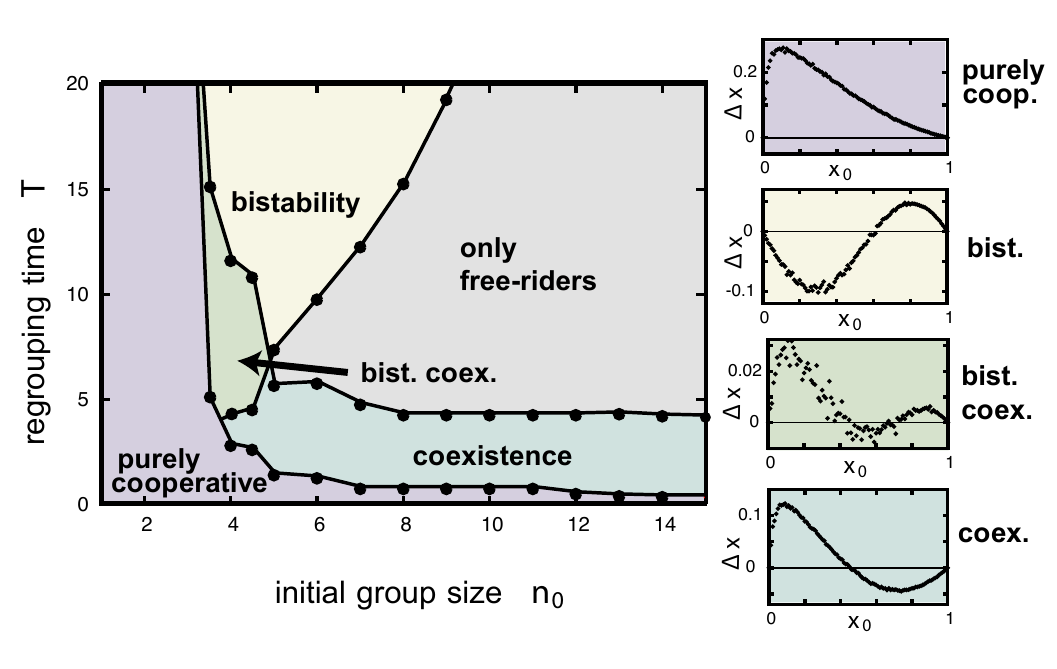}
\caption{{\bf Life cycle phase diagram.}
Fate of the metapopulation under life cycle dynamics as a function of the size of the population bottleneck $n_0$ and the regrouping time $T_0$; filled black cricles are simulation results from Ref.~\cite{Melbinger:2015}.
For a single life cycle step, the initial cooperator fraction $x_0$ is mapped to a final cooperator fraction $x(T)$.
The resulting drift $\Delta x(x_0)$ is shown for four different scenarios (right panles): purely cooperative regime ($n_0=4$, $T=1.5$), bistable regime where the map exhibits an unstable fixed point at $0<x^*<1$  ($n_0=5$, $T=20$), bistable coexistence where the map has both a stable and an unstable fixed point ($n_0=4$, $T=5.5$), and coexistence regime with a single stable fixed point at some intermediate value $0<x^*<1$ ($n_0=6$, $T=1.8$). 
Adapated from Ref.~\cite{Melbinger:2015}.
\label{fig:phases}
}
\end{figure}

Can such a transient increase in cooperator fraction lead to the evolution and maintenance of cooperation? 
This question has recently been addressed for the idealised restructuring scenario (life cycle) discussed in the introduction to this section~\cite{Cremer_Melbinger:2011, Melbinger:2015}; see the illustration in Fig.~\ref{fig:cartoon}. These studies have shown how the population structure depends on two key parameters, the initial population size $n_0$ and the time between repeatedly regrouping the subpopulations (regrouping time $T$).
The main insight gained from these theoretical studies is that there are two distinct mechanisms that can promote cooperation~\cite{Cremer_Melbinger:2011, Melbinger:2015}. 
(i) In the \emph{group-fixation mechanism}, the main effect is that during population bottlenecks with a very small number of individuals a significant fraction of subpopulations might fixate to purely cooperative colonies. 
(ii) In the \emph{group-growth mechanism}, cooperation is favored since subpopulations with a higher fraction of cooperators grow comparably faster and thereby compensate for the selection advantage of free-riders. In the following we review these mechanisms in more detail and discuss the ensuing `phase diagram' shown in Fig.~\ref{fig:phases}.

\paragraph{Group-fixation mechanism} 

If the subpopulations (groups) evolve separately for times much larger that the growth rate  ($T\gg 1$, time-scale growth-rate), it is likely that all of them reach a stationary state, where the population has fixated to either cooperators or free-riders. 
Since the corresponding groups sizes are then given by $(1+p)(1-s)K$ and $K$, respectively, the global fraction of cooperators will change from its initial value $x_0$ to~\cite{Cremer_Melbinger:2011, Melbinger:2015} 
\begin{align}
	x' (x_0)=
	\frac{(1+p)(1-s)P_C}{(1+p)(1-s)P_C+(1-P_C)} 
	\, .
\label{eq:xprime}
\end{align}
Here $P_C$ denotes the probability that a group after assortment consists of cooperators only~\cite{Melbinger:2015}. 
This equation can be read as an iterative nonlinear map for the cooperator fraction generated by the interplay between group evolution and regrouping. 
As $P_C$ generically increases with the initial fraction of cooperators, there is an unstable fixed point $x_u^*$, which implies bistability in the composition of the population as shown in Fig.~\ref{fig:phases}.
The final composition of the population depends on the initial cooperator fraction $x_0$. While for values above a threshold fraction $x_u^*$ the final population consists of cooperators only, cooperators become extinct in the long run if $x_0<x_u^*$.

\paragraph{Group-growth mechanism} Within groups the cooperator fraction always declines due to the selective advantage of cooperators. However, groups containing a larger fraction of cooperators grow stronger. This implies that more cooperative groups (subpopulations) contribute with a larger weight to the metapopulation average, an effect that is strongest during the initial growth phase of each subpopulation. The relative strength of selection advantage of free-riders within groups, $s$, and the growth advantage of more cooperative groups, $p$, is the decisive factor that determines whether this mechanism is strong enough to compensate for the selection disadvantage of cooperators~\cite{Cremer_Melbinger:2011, Melbinger:2015}. 
This group growth mechanism leads to a stable cooperator fraction (fixed point) $x^*_S$ for life cycles with repeated regrouping. 
A key insight is that the group-growth mechanism acts for much stronger population bottlenecks, $n_0$, than the group-fixation mechanism.
The underlying reason is that the mechanism relies on variance in group composition, but not on the existence of purely cooperative groups.
On the other hand, the group-growth mechanism only works for short regrouping times because it is caused by the initial growth advantage of more cooperative groups.

\paragraph{Phase diagram} The two key parameters of the life cycle model are the regrouping time $T$, and the initial group size $n_0$ (size of the population bottleneck). As we have just discussed, these two parameters determine which of the two mechanisms, group-growth or group-fixation, are dominant. As shown in Fig.~\ref{fig:phases}, repeated regrouping leads to five distinct types of dynamics for the changes $\Delta x =x(T)-x_0$ in the cooperator fraction. As discussed in more detail in Ref.~\cite{Melbinger:2015}, there is a bistable regime and a stable coexistence regime for parameters $T$ and $n_0$ where the group-fixation and the group-growth mechanism dominates, respectively. Moreover, while at small groups sizes the metapopulation becomes purely cooperative (due to group fixation events), free-riders completely take over the metapopulation for large $n_0$ and large regrouping times (due to their selection advantage).

Overall these considerations show that the combination of reoccurring population bottlenecks and population growth can stabilize cooperative traits. The continuous restructuring of the population can also allow the evolution of cooperative behavior when starting with a low fraction of cooperative individuals (in the extreme case a single mutant)~\cite{Melbinger:2015}. Details of public good synthesis and utilization affect growth behavior and costs. This can shift the time-scales of restructuring ($T$) and bottleneck sizes ($N$) required to stabilize cooperation. 

\section{The public good dilemma in spatially extended systems}
\label{sec:spatially_extended_systems}

Up to now, we mostly discussed scenarios where local (sub)-populations are well-mixed. However, the explicit consideration of space is an important aspect of microbial life as many microbes on our planet occur in the form of dense microbial communities like colonies and biofilms~\cite{Nadell:2016, Flemming:2019}, quite distinct from a well-mixed scenario. Even more, spatial extension and arrangement can strongly affect evolutionary dynamics~\cite{Yanni:2019ip}.
In this section, we thus give a short overview of experimental and theoretical work on the public good dilemma in spatially extended systems.  

\subsection{Modeling cooperation in spatially extended systems}
\label{sec:modelingspatialextendedsystems}

As we discussed in Section~\ref{sec:factors_cooperationinmicrobialcooperation}, Hamilton already argued that spatial clustering of strains producing a public good (cooperators) may support cooperation populations~\cite{Hamilton:1964}. 
In spatial clusters, cooperators can preferentially interact with each other and thereby are less likely to be exploited by free-riders. 
Theoretically, this idea has been studied early on by Nowak, Bonhoeffer and May in the context of the \emph{spatial prisoner's dilemma game}~\cite{Nowak_Bonhoeffer:1994}. 
In this variant of the dilemma, individuals are arranged on a lattice and only interact with their nearest neighbors. 
This class of conceptual theoretical models has been explored quite thoroughly using a variety of deterministic and stochastic interaction rules. 
These studies confirm that formation of cooperative clusters can promote cooperation in such setups, see e.g.\/ Refs.~\cite{Nowak_Bonhoeffer:1994, Nakamuru:1997, Langer:2008, Szabo:2009, Helbing:2009, Fu:2010, Szolnoki:2009, Liu:2012, Szolnoki:2009a, AlonsoSanz:2009, Wang:2012}. 
Theoretical modeling was also extended to more complex structures, beyond simple lattices, including interactions in networks, see e.g.\/ Refs.~\cite{Ohtsuki:2006, vanBaalen:1998, Ifti:2004, Santos:2005, Abramson:2001, Lieberman:2005, Szabo, Vukov:2006}. 
In fact, in many of these systems, stable cooperation was obtained. 
However, at the same time, the stability of cooperation often depended on the exact details of the underlying dynamical implementation of the cellular automatons and dynamical systems studied.

In the context of microbes producing a diffusible public good, many specific assumptions of these various models are probably unrealistic. 
Nevertheless, the clustering of cooperation might lead to stable cooperation if one or both of the following requirements are met (similarly to what we discussed in Section~\ref{sec:factors_cooperationinmicrobialcooperation}): 
There is some positive assortment (spatial segregation) of cooperating individuals~\cite{Fletcher_Doebeli:2006} stably maintained over time, and the diffusible public good does not spread out evenly in the whole population, but preferentially remains in producer-rich areas~\cite{Borenstein2013}. 
Greater sharing of public goods leads in general to a reduction of cooperation~\cite{Menon:2015bu}.

\subsection{Experimental studies with microbes}

The dynamics of spatially extended populations is quite complex, even for well-controlled and idealized experimental setups.
There are various reasons for this complexity: 
(i) Within the dense communities, strong consumption and synthesis leads to strongly heterogeneous profiles of metabolites and waste-products~\cite{Tolker-Nielsen:2000, Pagaling:2017, Flemming:2019, Warren:2019cl}. 
(ii) Similar as in well-mixed systems, there is a regulatory network that controls the production of public goods and other exoproducts like, for instance, quorum sensing molecules; see discussion in Section~\ref{sec:pyoverdine}. 
(iii) This leads to an intricate coupling and feedback between the environment shaped by the production of these products and the microbes growing in this self-shaped environment, sometimes referred to as eco-evolutionary feedback~\cite{Sanchez_Gore:2013, Bauer_Knebel:2017}. 
Differential motility and growth of microbes, growth of the microbial colony as a whole (range expansion), and externally imposed spatial and temporal variations of the environment are some other factors that in combination with regulatory feedback lead to a complex intertwined dynamics.  
Nevertheless, experiments performed under controlled laboratory conditions have led to considerable insights into cooperation within such communities. 

\paragraph{Range expansion}
Range expansion refers to the growth of a population into a previously unoccupied territory and is supposed to occur in response to changes in the environment. 
This type of population growth can have a major effect on the spatial arrangement and the composition of the population~\cite{Hallatschek:2007, Korolev:2010, Hallatschek:2010, Kuhr:2011, Korolev:2012hx, Korolev:2013bc, SenDatta:2013gj, Fusco:2016hga, weber_interface_2014, Reiter:2014, Gralka:2019kw}. 
For example, monoclonal sectoring patterns that arise as a consequence of random genetic drift  drives population differentiation along the expanding fronts of bacterial colonies~\cite{Hallatschek:2007, Korolev:2010}. 
This segregation process strongly favors cooperation in a yeast population emulating a prisoner's dilemma game~\cite{VanDyken:2013fg}, affects mutualism in yeast colonies~\cite{Muller:2014, Momeni:2013ei, Momeni:2013dw,Menon:2015bu}, and promotes biodiversity in an \textit{E. coli} system with cyclic dominance between strains~\cite{weber_interface_2014}. 
Moreover, genetic drift at expanding fronts can also lead to full fixation of certain strains~\cite{Excoffier:2009, Korolev:2010}.
Using a yeast model system~\cite{Gore:2009} and a spatially extended setup that emulates a stepping stone model~\cite{Kimura_Weiss:1964, Korolev:2010, Korolev:2011, Korolev:2013bc}, it was found that cooperation can be maintained through enrichment of cooperators at the front~\cite{SenDatta:2013gj}.Recently, it was shown that the inevitable mechanical interactions between cells can also significantly affect the fitness of cells at the expanding edge of yeast colonies~\cite{Kayser:2018}.

\paragraph{Biofilms}
Biofilms have become a particular focus of recent research~\cite{Kreft:2004gh, Drescher:2014, Nadell:2016, Flemming:2019}, not only because of their high abundance in nature and their relevance in understanding the role of cooperativity and competitive cell-cell interactions~\cite{Kreft:2004gh, Nadell:2016} but also as model systems for studying the evolution of multi-species communities in general, including the division of labor~\cite{Dragos:2018} and the interaction with phages, their important predators~\cite{Vidakovic:2018ij}. 
From experimental studies on biofilms a multitude of effects, often physical by nature, have been identified to contribute to the maintenance of cooperation.
Biofilm thickness has been shown to limit the dispersal of public good confining it to producer-rich regions, and thereby promoting cooperation~\cite{Drescher:2014}. 
In microcolonies of \textit{P. aeruginosa} growing on solid substrates, there is evidence that pyoverdine is directly exchanged between neighboring cells rather than by global diffusion~\cite{Julou:2013}.
In experimental populations of \textit{P. fluorescens}, cooperating groups are formed by over-production of an adhesive polymer, which causes the interests of individuals to align with those of the group~\cite{Rainey:2003}. 
This is by far not an exhaustive list of such physical mechanisms; others are discussed in recent review articles~\cite{Hibbing:2010, Archetti-Review:2012, Celiker2013}.

\paragraph{Artificial spatially extended microhabitats}
There have also been advances in designing artificial spatially extended habitats using microfluidic devices which allow to experimentally control the size of communities (patches) and how strongly these couple to neighboring communities~\cite{Keymer:2006, Keymer:2008, Hol_Keymer:2013}. 
Using an \textit{E. coli} community that emulates a `social' interaction between cooperators and free-riders~\cite{Hol_Keymer:2013}, coexistence of both strains was observed if there is some kind of spatial organization.
For the future, it would be interesting to have experimental studies along theses lines. 
This may allow to learn how additional effects of phenotypic heterogeneity and plasticity, the dynamics and regulation of public goods, and phenotypic heterogeneity in the bacteria's mobility favor or disfavor the emergence of cooperative behavior in microbial model populations. 

\subsection{The role of bacterial motility in spatially extended systems}

For most of the experimental conditions employed in the studies discussed in the previous section (with the exception of artificial microhabitats), active motility of the microbes can be disregarded. However, many microbes have sophisticated mechanisms to actively move, and this motility can strongly challenge the evolutionary stability of cooperation. It is thus important, to consider the interplay of motility and the emergence of cooperative behavior.

Already undirected movements (e.g. diffusion), can foster the invasion of free-riders into cooperative clusters and thereby counteracts the evolution of cooperation.  
How fragile this spatially promoted maintenance of cooperative behavior can be has been illustrated by a theoretical study that takes into account two essential features of microbial populations: motile behavior and its interplay with the processes of fitness collection and selection that happen on vastly different time scales~\cite{Gelimson:2013}. 
While stable cooperation emerges without motile behavior, as has been reported for the spatial prisoner's dilemma~\cite{Nowak_Bonhoeffer:1994} (see also Section~\ref{sec:modelingspatialextendedsystems}), it is found that even small diffusive motility strongly restricts cooperation since it enables non-cooperative individuals to invade cooperative clusters. 
This suggests that in many biological scenarios, the spatial clustering of individuals cannot explain stable cooperation but additional mechanisms are necessary for spatial structure to promote the evolution of cooperation. 
For example, it has recently been shown that delayed adjustment to changes in the environments can affect the long-term behaviour of microbial communities and promote robust  cooperation~\cite{Bauer_Frey_PRL:2018, Bauer_Frey_PRE:2018, Bauer_Frey_EPL:2018}.

Active motility is an essential part of biological reality. 
Many different microbes are motile, showing not only undirected movement, but also highly directed movement along sensed gradients (chemotaxis)~\cite{Spormann:1999il, Baym:2016hx, Fraebel:2017gw, Liu:51NDZHe4, Cremer:1LtZBIVB, Jeckel:2019}
Often, swimming modes include collective phenomena like swarming and the effective expansion of populations into new habitats. 
In fact, collective expansion by random movement~\cite{Wei:2011ca} and even more so collective expansion by directed movements along population driven gradients~\cite{Cremer:1LtZBIVB} are efficient mechanisms to ensure fast growth and the successful competition for nutrients~\cite{Cremer:1LtZBIVB, Liu:51NDZHe4}. 
Thus, if both undirected motility threatens cooperation and motility is part of biological reality contributing to fitness, this raises the question how fundamental spatial clustering as a mechanism to explain cooperation really is. 
Again, the answer is expected to depend strongly on the specific strains and ecological conditions one considers. 
In many situations, where public goods are involved, swimming might still be negligible. 
In other situations, however, strong swimming and public good production might go hand in hand and cooperation is stable because of the interplay between growth and competition in restructuring populations.
Further experimental studies are needed to investigate these aspects.

\section{Conclusions and Outlook} 
\label{sec:conclusions_outlook}

It seems that by asking how cooperation may have emerged and is maintained in microbial populations we have arrived at more questions than answers. Indeed, as we have emphasized throughout this review article, there is no unique answer to this challenging question. In the following, we will give a concise summary to the answers we found for populations with a given population structure, and close with a section highlighting some of the future challenges in the field.  

\subsection{Conclusions}
\label{sec:conclusions_outlook_conclusions}

In this review we considered the dilemma of cooperation in microbial populations, and asked what biological and environmental factors can promote the emergence and maintenance of cooperation in structured populations. Specifically, we focused on population structures (life cycles) with recurring population bottlenecks and strong episodes of population growth. As we discussed, all these aspects are typical features of a broad range of microbial species and ecological settings. The (partial) answer to the dilemma of cooperation obtained from these studies is that regular dispersal leading to population bottlenecks and subsequent population growth can promote cooperation. In that sense, the mechanism is diametrically opposed to Hamilton's suggestion that limited dispersal can facilitate cooperation~\cite{Hamilton:1964}. A closer look though reveals that in these structured populations it is actually the interplay between limited dispersal and strong dispersal resolving the dilemma. 
During the growth phase after population bottlenecks, confinement to a sub-population (limited dispersal) ensures that public goods shared between cooperating (producing) individuals benefits these sub-populations such that they have a growth advantage against sub-populations containing less cooperative individuals. Without dispersal, however, in each of these sub-populations non-producing individuals would in the long run be better off and become the dominant species. The only way to avoid such `take over' is recurrent dispersal into smaller groups. 

While the above insights can already be gained by analysing abstract theoretical models using a framework that combines evolutionary game theory with population dynamics~\cite{Melbinger:2010, Cremer_Melbinger:2011, CremerPRE:2011, Melbinger:2015}, a quantitative understanding of a specific biological system requires an in-depth analysis of the synthesis and regulation of the public good. We illustrated this for an experimental model population of \textit{Pseudomonas putida}. We showed how a combined approach using theory and experiment provides insights into how the interplay between selection pressure, growth advantage of more cooperative sub-populations, and demographic noise affect the dynamics of the entire population.  Chemical and biological properties of the public good --- such as its stability and a dose-dependent synthesis and benefit --- were identified as important factors leading to the emergence and maintenance of public-good-providing traits.

\subsection{Future challenges}
While theoretical studies have revealed some general principles that facilitate cooperation, the results discussed in this review also clearly show that there is a lack of understanding in the cellular and environmental factors that regulate public goods.
Therefore, future research should avoid preconceived notions about the interactions between microbes and instead rather focus on an analysis and understanding of the specific biological and ecological conditions.

\paragraph{Biochemical regulation of public goods} 
Studying the public good pyoverdine of \textit{Pseudomonas} populations in controlled environments, we have seen that biochemical details --- such as public good recycling and accumulation --- affects growth and long-term population dynamics.
Accumulation, in particular, has far-reaching consequences on the possibility of sustaining the observed increase in producer fraction.
In fact, even if producers could ``cut their losses'' and stop production when some threshold concentration is reached, the benefit from more abundant pyoverdine would eventually vanish.
The public good, in fact, gradually accumulates also in populations with few producers, allowing them to eventually reach the same size and pyoverdine concentrations of more producing ones.
Therefore, any population-level advantage from such an accumulating public good is bound to vanish in time, and the regulation of pyroverdine production and its advantage and costs have to be considered within the ecological context cells live. 

Perhaps the most interesting avenue of further research is, then, a rigorous modeling of the regulatory network, showing in detail the impact of pyoverdine and iron concentrations on synthesis, but also on cell metabolism. In this review, in fact, we discussed experimental results obtained with a constitutive producer strain, but wild-type pseudomonads (for example \textit{P. putida} KT2440) strictly regulate production.
This complex gene network involves central metabolic regulators and even genes of unknown function \cite{swingle_characterization_2008, cornelis_iron_2009}.
Pyoverdine concentrations, thus, could affect growth rates beyond iron availability, potentially providing further benefits to producer strains. 
In addition, it would be interesting to consider bacterial iron acquisition systems without siderophore recycling. 
For example, in \textit{Escherichia coli} the siderophore enterobactin is degraded after being used for iron uptake~\cite{Raymon:2003}.


In this review we focused on controlled environments as provided in simple laboratory experiments. While a good starting point, it is also clear that we eventually have to understand evolution and the role of public goods within native environments bearing much higher levels of complexity, including multiple species, stress factors like pH levels or salt concentrations detrimental to growth, and a complex mixture of nutrients limiting growth in various ways. As with exoproducts microbes release, like toxins or exopolysacherides, one has to critically asses the assumed roles of public goods. Beyond some cooperative interaction, the public goods discussed in this review might be involved in many other interactions and a simple focus on cooperation might be misleading. 

\paragraph{Resource abundance and microbial growth}
Towards a more complete picture of public goods and their role in population growth and evolution, many more investigtions are required. This will particularly require to consider specific ecological scenarios.  For example, future investigations could study \emph{Pseudomonads} within a certain soil type, or certain gut bacteria within a certain niche of the gut. Specific quantities to consider include the energy supply which drives microbial growth in the specific ecological scenario, the time and length scales involved in nutrient supply and depletion, and the emerging population structure related to these factors. For example, one may ask what is driving population structure and growth in soil or within the human gut?

\paragraph{Metabolic exchanges and community dynamics}
The exchange of metabolites and their effect on cell growth and survival is another  important factor to consider. 
Metabolic exchanges depend on the specific ecological scenario one is studying and can also shift the benefits and costs associated with cooperative behavior.  More detailed investigations in the future should thus also consider specific metabolic interactions and integrate our current understanding of metabolite exchanges and microbial community assembly~\cite{Friedman:2017bz, Chacon:2018, Warren:2019cl}. 

\paragraph{Quorum sensing and cell-to-cell signaling}
The exchange of signaling molecules is another important factor which can strongly interfere with the emergence and stability of cooperative behavior. Cellular signaling can tightly couple the expression of public goods to community sizes (quorum sensing) and environmental conditions (environmental sensing). Signaling via autoinducers is known to be involved in the regulation of various pulic goods, including different types of pyoverdines. Such feedbacks can strongly affect the costs and beneficial aspects of public goods synthesis~\cite{Heilmann:2015dz, Gupta:2013cy, Schuster:2017}.

\paragraph{Environmental noise and catastrophes}
Further, the fate of microbial populations is  affected by a number of environmental factors, such as the presence of toxins, temperature, light, pH, or phages~\cite{Morley:1983, Fux:2005}. 
Changes of these factors can be fast, leading to dramatic shifts of environmental conditions. Often, microbes themselves are triggering such shifts.  Consider for example the lysogenic and lytic spreading of phages~\cite{Vasse:2015, Moebius:2015, Simmons:2018iv}, or the fall of local pH values,  caused by acidic fermentation products~\cite{Cremer:2017bm, Arnoldini:2018cc, Ratzke:2018, Ratzke:2018b}. Conversely, such sudden environmental changes can drastically effect the fitness landscape, strongly selecting for genes increasing viability and survival.  This again changes the costs and benefits of puplic good synthesis. 
The dramatic shifts also affect the population size and can - similarly as dispersal - lead to population bottlenecks where effects from demographic noise are particularly pronounced. 
Recent studies using conceptual theoretical models have addressed how environmental fluctuations affect the composition of populations, consisting for example of a fast growing strain competing with a slow growing (cooperating) species~\cite{Melbinger_Vergassola:2015,Wienand_Frey_Mobilia:2017, Wienand_Frey_Mobilia:2018}. As with earlier conceptual studies,
the future challenge will be to connect these with specific microbial model systems under controlled - but hopefully realistic environmental conditions. 

\paragraph{Multi-species communities}

Microbial communities often consist of many different species, adding another layer of complexity. In this review we focused on closely related strains, varying only in their capability to produce public goods. Genetic and phenotypic differences in real microbial communities can however be very strong. For example, sequencing studies have shown that specific environments are often occupied by strains from different bacterial phyla and geni~\cite{Hugerth:2017, TheEarthMicrobiomeProjectConsortium:2017}. A dramatic example is the gut microbiota of vertebrates, which consists of up to hundreds of different bacterial species, varying vastly over space and time~\cite{costello_application_2012,LloydPrice:2017}.  
While a strong functional redundancy of genes across different species is typically  observed~\cite{Louca:2018}, each species can show distinct growth and survival phenotypes.  The species richness and functional redundancy can have strong impacts on community function and thus also on the evolutionary stability of cooperative behavior. For example, cross-feeding within these complex communities involves the exchange of many different metabolites~\cite{Nemergut:2013}, affecting the cellular resources allocated to different cellular processes and thus also the benefits, costs and selection pressures of cooperative traits.

\paragraph{Spatially extended systems and biofilms}
Going beyond well-mixed systems leads to further challenges in understanding the population dynamics of microbial communities. 
For example, physical factors might further promote cooperation. Recent theoretical suggestions include differential adhesion between microbes~\cite{Garcia:2014}, the formation of cooperator aggregates promoting preferred access to public goods~\cite{olejarz_jtb_2014}, different types of fluid flow~\cite{Drescher:2014, Groselj:2015}, and biofilm formation~\cite{Nadell:2016, Flemming:2019}.  Increased phage resistance by biofilm formation is another important direction to go~\cite{Vidakovic:2018ij}. More attention should also be given to the role of phenotypic heterogeneity and plasticity which, as discussed above, is an important feature of pyoverdine-producers such as \textit{P. putida}. \\
\\
Addressing the challenges described here requires comprehensive interdisciplinary research approaches which  combine the different ecological and biological scales involved. This includes particularly the environmental conditions supporting the growth of microbial communities, evolution and the ecological interactions within these communities, and the physiological characterization of the single species involved. We believe that modeling will be an essential part of this research path as it allows to investigate the interplay of these different levels at play.

\section{Acknowledgements}
We would like to thank Felix Becker, Johannes Knebel, Matthias Lechner, and Paul Rainey for insightful and stimulating discussions. Further, we would like to thank Kirill Korolev, Wolfram M\"obius and the reviewers for their insightful feedback. This work was financially supported by the Deutsche Forschungsgemeinschaft (DFG) through projects JU 333/5-1,2 and FR 850/11-1,2 within the grant on ``\emph{Phenotypic heterogeneity and sociobology of bacterial populations}" (SPP1617). 

\section{References}


\cleardoublepage
\newpage

\appendix 
\renewcommand*{\thesection}{\Alph{section}}

\section{Relation between Price and replicator equation}
\label{app:price_replicator}

As mentioned in the main text, the Price equation can be mapped to the replicator equation. This mapping is shown here. We start from the adjusted replicator equation
\begin{equation}
    \dot x_k 
    = 
    \frac{f_k-\bar f}{\bar f} 
    \, x_k \, (1-x_k)\, ,
\label{eq:rep}
\end{equation}
where $f_k$ is the fitness of a certain trait $k$, $x$ is the relative abundance of just this trait and  $\bar f=\sum_k x_kf_k$ is the average fitness in the population.
In contrast to the replicator equation, where the fractions of certain traits, $x_k$ are considered,  the  Price equation analyzes the values of the traits, $z_i$. 
To achieve a mapping, the more general quantity, $z_i$,  has to be chosen appropriately: The trait $z_i$ now marks the belonging of an individual, $i$, to a certain species,  $k$. Each species, $k$, can be distinguished by its typical value of the trait $z_k$. We therefore define new traits whose values  are $\tilde z_i^{(k)}=1$ if the individual $i$ is of type $k$ and  $\tilde z_i^{(k)}=0$ if it belongs to any other species. This can be summarized to $\tilde z_i^{(k)}=\delta_{z_i,z_k}$. The fraction of species $k$  is then given by  $x_k=\langle\tilde z_i ^{(k)}\rangle=\sum_i h_i \delta_{z_i,z_k}$. The growth factor, $w_i$, in the Price equation corresponds to the fitness of an individual and therefore solely depends on the species the considered individual belongs to. Therefore, it is given by $w_i=\sum_l\delta_{z_i,z_l}f_l$. For example  an individual, $i$, belonging to species $k$ has the growth factor $w_i=f_k$. The average growth factors is then given by the average fitness in the population, $\langle w_i\rangle=\bar f$. As mutations are not included, $\tilde z_i^{(k)}$ does not change and $\Delta \tilde z_i^{(k)}=0$ holds. Taken together, this yields the following modification of the Price equation~\eqref{eq:Price},
\begin{equation}
    \Delta x_k \, \bar f
    =
    \sum_i h_i \, 
    \tilde z_i^{(k)}w_i 
    - 
    \bar f x_k \, . 
\label{eq:Pricetorep}
\end{equation}
The term $\sum_i h_i \tilde z_i^{(k)} w_i=\sum_{i,l} h_i \delta_{z_i,z_k}\delta_{z_i,z_l} f_l$ is nonzero only if the considered species is of type $k$. Then its fitness is always given by $f_k$. Therefore, one finds that $\sum_i h_i \tilde z_i^{(k)} w_i=f_k\sum_i  h_i \delta_{ z_i,z_k}= f_kx_k$ holds and Eq.~\eqref{eq:Pricetorep} simplifies to,
\begin{equation}
    \Delta x _k\bar f
    =
    (f_k-\bar f) \, x_k
    \, . 
\label{eq:Pricetorepfin}
\end{equation}
This expression is equivalent to the adjusted replicator equation for discrete time steps. Performing the limit $\Delta t\rightarrow 0$ then gives Eq.~\eqref{eq:rep}.

\section{Selection on two levels and Hamilton's rule}
\label{app:hamiltonsrule}

To mathematically analyse the situation, we consider the Price equation as introduced in Section~\ref{sec:price_equation}. As we specifically consider two levels of selection here, we also formulate the Price equation with two levels. All quantities previously introduced in Sec.~\ref{sec:price_equation} now come in two forms, for the intra- and inter-group level~\footnote{This analysis follow~\cite{Melbinger:2011thesis,Cremer:2011thesis}}. Let us start with the intra-group level: Each individual therein is classified by its trait $z_{i,\alpha}\in\{0,1\}$ where 0 corresponds to  a free-rider and 1 to a cooperator. The index $i$ specifies the individual and $\alpha$ its group. The factor $h_{i,\alpha}$ corresponds to a trait's abundance. Summing them up leads to the average trait of a group  $Z_\alpha= \langle z_{i,\alpha}\rangle_\alpha=\sum_i z_{i,\alpha}h_{i,\alpha}$. The reproduction success of an individual in the considered time interval $\Delta t$ is given by the fitness factor $w_{i,\alpha}$.

Thus, the Price equation, describing the change of the average trait in each group is given by:
\begin{equation}
\Delta \langle z_{i,\alpha}\rangle_\alpha\langle w_{i,\alpha}\rangle_\alpha=\text{Cov}[z_iw_i]_\alpha.
\end{equation}

In this equation, the term $\langle \Delta z_{i,\alpha}w_{i,\alpha}\rangle_\alpha$ is not present as mutations towards different phenotypes are not considered and $ \Delta z_{i,\alpha}=0$ holds. For the cooperation scenario, cooperators have a fitness disadvantage within groups. Thus, the covariance between individual trait value and its fitness is negative. As a consequence, the average trait value within each group, independently of its internal composition. An example of this is shown in Fig.~\ref{fig:two_level}(c).

Next, let us consider the upper level, comparing different groups. Each group has the trait $Z_\alpha$, depending on the individuals within this group, $Z_\alpha=\langle z_{i,\alpha}\rangle_\alpha$. At the same time, competition between the groups (inter-group selection) is described by the Price equation:
\begin{eqnarray}
\langle\Delta  Z_\alpha \rangle\langle W_\alpha\rangle=& \text{Cov}[Z_\alpha W_\alpha ]+\langle\Delta Z_\alpha W_\alpha\rangle. 
\label{eq:inter}
\end{eqnarray}
Here, the  fitness factor $W_\alpha$ describes the relative success of different group.
In Table.~\ref{table:Price} the corresponding terms on both levels are summarized. Note, that there are two kinds of averages, the one within a group summing over all individuals, $\langle x_{i,\alpha}\rangle_\alpha$, and the inter group average summing over all groups, $\langle X_\alpha\rangle$.

\begin{table}
\centering
\begin{tabular}{|c|c|c|c|c|c|c|}
\hline
Level & Ind. &Abu.& Average& Small Entity & Large Entity & Gr. F. \\ \hline\hline

Inter & $\alpha$ &$H_{\alpha}$& $\langle X_\alpha\rangle= \sum_\alpha X_{\alpha}H_{\alpha}$ & Group $Z_{\alpha}$ &  Set of groups $\langle Z_\alpha \rangle$& $W_\alpha$ \\ \hline
Intra & $i$ &$h_{i,\alpha}$ &$\langle x_{i,\alpha}\rangle_\alpha=\sum_i x_{i,\alpha}h_{i,\alpha}$& Individual $z_{i,\alpha}$ & Group $\langle z_{i,\alpha}\rangle_\alpha=Z_\alpha$& $w_{i,\alpha}$ \\ 
\hline
\end{tabular}

\caption{Comparison of the different quantities and averages to describe the simultaneous selection on the inter- and intra-group level.  From left to right, the summation index, the abundance, the average, the smaller and the larger entity and the growth factors are shown
\label{table:Price}}
\end{table}

Mathematically, an increase in the global level of cooperators corresponds to $\langle\Delta  Z_\alpha\rangle>0$. Employing the price equations on both levels, this inequality can be somewhat simplified as we mathematically discuss in Appendix~\ref{app:hamiltonsrule}. The inequality obtains a form:
\begin{eqnarray}
\mathcal{B}\mathcal{R}>\mathcal{C}.\label{eq:Hamiltonsrule}
\end{eqnarray}
If this inequality is true, than the inter-group level dominates, and the global level of cooperation increases. Otherwise, the global level of cooperation decreases. 

To simplify the condition for an increase of global population levels consider the inequality $\langle\Delta  Z_\alpha\rangle>0$. Importantly, average level of cooperation within each group depends on the individual cells. Let us thus start with the price equation describing traits within groups, Equation~\ref{eq:intra}. By multiplying this expression with $H_\alpha$ and summing over all groups, $\alpha$, it transforms to,
\begin{eqnarray}
 \sum_\alpha H_\alpha\Delta \bar z_\alpha\bar w_\alpha&=&\sum_\alpha H_\alpha \text{Cov}[z_iw_i]_\alpha \nonumber\\
 \langle\Delta Z_\alpha W_\alpha\rangle&=&\langle \text{Cov}[z_iw_i]_\alpha\rangle.
\label{eq:intra}
\end{eqnarray}

Combining Eqs.~\eqref{eq:inter} and \eqref{eq:intra} leads to the following condition for the regime of stable cooperation,
\begin{equation}
\text{Cov}[Z_\alpha,W_\alpha]+\langle \text{Cov}[z_iw_i]_\alpha\rangle>0.
\end{equation}
Now, the identity $\text{Cov}[A_\alpha B_\alpha]_\alpha=K(A_\alpha,B_\alpha)\text{Var}[A_\alpha]$ (and accordingly on the intra-group level $\text{Cov}[a_{i,\alpha} b_{i,\alpha}]_\alpha=k_\alpha(a_{i,\alpha},b_{i,\alpha})\text{Var}[a_{i,\alpha}]_\alpha$) following from linear regression can be used to further simplify the inequality,
\begin{equation}
K(Z_\alpha,W_\alpha)\,\text{Var}[Z_\alpha]+\langle k_\alpha(z_{i,\alpha},w_{i,\alpha})\,\text{Var}[z_{i,\alpha}]_\alpha\rangle>0.
\label{eq:lin_reg}
\end{equation}
The factor $ k_\alpha(z_{i,\alpha},w_{i,\alpha})$ corresponds to the disadvantage of cooperators within each group. If this disadvantage does not depend on the group $\alpha$, which is for example the case for public good producing bacteria whose metabolic disadvantage is independent of the group composition, Eq.~\eqref{eq:lin_reg} can be further simplified:
\begin{equation}
\frac{\text{Var}[Z_\alpha]}{\langle\text{Var}[z_{i,\alpha}]_\alpha\rangle} \, K(Z_\alpha,W_\alpha)>-k_\alpha(z_{i,\alpha},w_{i,\alpha}).
\end{equation}
This expression is a general form of Hamilton's rule, $\mathcal{R}\cdot\mathcal{B}>\mathcal{C}$ with the relatedness, benefit and cost functions given as 
\begin{eqnarray}
    \mathcal{R} &=& {\text{Var}[Z_\alpha]}/{\langle\text{Var}[z_{i,\alpha}]_\alpha\rangle} \, \\ 
    \mathcal{B}
    &=& K(Z_\alpha,W_\alpha) \, ,\\ 
    \mathcal{C}
    &=& -k_\alpha(z_{i,\alpha},w_{i,\alpha}) \, .
\end{eqnarray}

The significance of this rule is briefly discussed in Section~\ref{sec:note_hamiltonsrul}. Here we further note that the notion of a group structure can mathematically be used in a very general sense and does not rely on the strict spatial separation of groups. Further, many studies have also investigated very different structures, including network-structures or interacting with neighbouring individuals on a lattice; see also the discussion in Section~\ref{sec:spatially_extended_systems} were we discuss spatially extended systems were local sub-populations are not well mixed.

For completeness, we note that the Price equation has been extended to describe evolution across several levels~\cite{Hamilton:1975, Okasha}. The framework is hence often called the framework of \emph{multi-level selection} and such approaches have been discussed in the context of major transitions~\cite{Keller:1999, Okasha}.

\end{document}